\documentclass[a4paper,11pt]{article}
\usepackage{jheppub} 
\usepackage{tikz}
\usepackage{ytableau} 
\usepackage{enumitem}
\usepackage[mathscr]{euscript}
\usepackage{tikz}
\usepackage{amsmath, amssymb}
\usepackage{booktabs}
\usepackage{caption}
\usepackage{hyperref} 

\newtheorem{example}{Example}

\def\beq#1\eeq{\begin{align}#1\end{align}}

\title{Bigraded Polynomials for the Cohomology of Wild Hitchin Systems}

\author[~]{Dan Xie}

\affiliation[a]{Department of Mathematics, Tsinghua University, Beijing, 100084, China}

\abstract{We introduce a bi-graded polynomial that encodes the cohomology groups of the wild Hitchin system of type~$A_{n-1}$, constructed using an irregular singularity (determined by an integer~$m$) and an arbitrary regular singularity~$f$. When the regular singularity is of the form~$f = [1, \ldots, 1]$, the bi-graded polynomial~$C_{m,n}(q,t)$ coincides with the bigraded rational parking function defined combinatorially, admitting a Schur expansion~$C_{m,n}(q,t) = \sum_\lambda f_\lambda(q,t) s_\lambda(x)$. For general~$f$, the polynomial takes the form~$C^f_{m,n}(q,t) = \sum_\lambda f_\lambda(q,t) K_{\lambda f}$, where~$K_{\lambda f}$ denotes the Kostka number. We conjecture that this bi-graded polynomial agrees with the one arising from the perverse filtration of the Hitchin fibration, or equivalently, from the weight filtration of the mixed Hodge structure from the character variety. We also give a description by using the geometry of affine Springer fiber.}

\begin{document} 
\maketitle
\flushbottom

\section{Introduction}

\begin{figure}[htbp]
    \centering
    \tikzset{every picture/.style={line width=0.75pt}} 
    \begin{tikzpicture}[x=0.45pt,y=0.45pt,yscale=-1,xscale=1]
        \draw   (332,192) .. controls (332,131.25) and (381.25,82) .. (442,82) .. controls (502.75,82) and (552,131.25) .. (552,192) .. controls (552,252.75) and (502.75,302) .. (442,302) .. controls (381.25,302) and (332,252.75) .. (332,192) -- cycle ;
        \draw   (436,117) -- (445,117) -- (445,126) -- (436,126) -- cycle ;
        \draw    (433,256.96) -- (444,267.96) ;
        \draw    (433,268) -- (444,255.96) ;
        \draw (466,113.4) node [anchor=north west][inner sep=0.75pt]    {$\Phi_{n,k} $};
        \draw (459,257.4) node [anchor=north west][inner sep=0.75pt]    {$f$};
    \end{tikzpicture}
    \caption{Schematic representation of the Riemann sphere with an irregular singularity $\Phi$ at $\infty$ and a regular singularity $f$ at $0$.}
    \label{fig:singularities}
\end{figure}
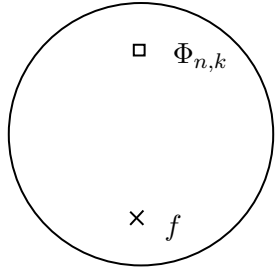

We study the cohomology groups of the $A_{n-1}$-type Hitchin moduli space defined on $\mathbb{P}^1$ with one irregular singularity and one regular singularity (see Figure~\ref{fig:singularities}). The irregular singularity is characterized by the Higgs field behavior \cite{Xie:2012hs}:
\begin{equation}\label{eq:irregular_singularity}
    \Phi_{n,k} = \frac{T}{z^{2+\frac{k}{n}}} + \cdots,
\end{equation}
where $k \geq -n$ is an integer coprime to $n$, and $T$ is a regular semi-simple element. The regular singularity $f$ corresponds to a nilpotent orbit of the $A_{n-1}$ Lie algebra, classified by a Young diagram $[n_1,\ldots, n_s]$ with $\sum_{i=1}^s n_i = n$.

The geometric data for the Hitchin moduli space $M_H(\Phi_{n,k}, f)$ is summarized in Table~\ref{tab:dimensions}. We include both the hyperkähler dimension of $M_H$ and the number of $\mathbb{C}^*$-fixed points, which directly relates to the total cohomology dimension \cite{goresky2006purity, goresky2004homology, varagnolo2009finite, oblomkov2016geometric, shan2023mirror, shan2024modularity}.

\begin{table}[htbp]
    \centering
    \caption{Dimensions and cohomology data for $M_H(\Phi_{n,k}, f)$. Here $m = n + k$, and $e_1, \ldots, e_{n-1}$ are the exponents of the regular singularity $f = [n_1, \ldots, n_s]$, given by $(0,1,\ldots,n_1-1,0,\ldots,n_2-1,\ldots,0,\ldots,n_s-1)/(0)$.
    $\dim \mathcal{O}_f$ is the hyperkahler dimension of the nilpotent orbit associated with $f$.}
    \label{tab:dimensions}
    \begin{tabular}{|c|c|}
         \hline
        $d = \dim M_H(\Phi_{n,k}, f)$ & $\sum \dim H^*(M_H,\mathbb{C})$ \\ \hline
        $\frac{(n-1)(m-1)}{2} - \dim \mathcal{O}_f$ & $\dfrac{(m-e_1)(m-e_2)\cdots(m-e_{n-1})}{(e_1+1)\cdots(e_{n-1}+1)}$ \\ \hline
    \end{tabular}
\end{table}

Our objective is to establish a bigraded structure on  cohomology groups of Hitchin's moduli space. The $q$-grading corresponds to cohomological degree, while the $t$-grading arises from two potential geometric sources:
\begin{itemize}
    \item[(a)] The perverse filtration $P_\bullet H^*$ induced by the Hitchin fibration $\pi: M_H \to B$, yielding the bi-graded polynomial \cite{de2010perverse}:
    \begin{equation}\label{eq:hodge_poly}
        H_P(q,t) = \sum_{i,k} \dim \mathrm{Gr}_{i}^P (H^k) q^i t^k.
    \end{equation}
    
    \item[(b)] The weight filtration $W_\bullet$ of the mixed Hodge structure on $H^*(M_H)$ ($M_H$ is regarded as character variety), giving the mixed Hodge polynomial:
    \begin{equation}\label{eq:hodge_poly}
        H_W(q,t) = \sum_{i,k} \dim \mathrm{Gr}_{2i}^W (H^k) q^i t^k.
    \end{equation}
\end{itemize}
The celebrated $P=W$ \cite{de2012topology, hausel2022p, maulik2024p} conjecture posits the equivalence of these filtrations for the $A$ type case with regular singularities, 
and we conjecture it is extended to the irregular case.

We introduce a related polynomial $C(q,t)$ defined by:
\begin{equation}\label{eq:C_poly}
    H(q,t) = q^d t^{2d} C(q, q^{-1} t^{-2}).
\end{equation}
where $d$ is the Hyperkhaler dimension of the Hitchin's moduli space.
Instead of computing the bi-graded polynomial by using either the perverse filtration or the mixed Hodge structure of the Hitchin's moduli space, we propose a combinatorial formula for $C_{m,n}(q,t)$:

\begin{enumerate}
    \item For the full regular singularity $f = [1,\ldots,1]$, $C_{m,n}(q,t)$ matches the rational parking function polynomial for $m \times n$ rectangles \cite{garsia1996remarkable, hikita2014affine, armstrong2016rational,gorsky2016affine}:
    \begin{equation}\label{eq:parking}
        \boxed{C_{m,n}(q,t) = \mathrm{PF}_{m,n}(q,t) = \sum_{P \in \mathrm{PF}(m,n)} t^{\mathrm{area}(P)} q^{\mathrm{dinv}(P)} F_{n,\mathrm{lDes}(P)}(x) = \sum_\lambda f_\lambda(q,t) s_\lambda.}
    \end{equation}
    The first sum ranges over labeled Dyck paths $P$ with statistics $\mathrm{area}(P)$, $\mathrm{dinv}(P)$, and descent set $\mathrm{lDes}(P) \subseteq \{1,\ldots,n-1\}$ determining a fundamental quasi-symmetric polynomial, while the second expression gives the Schur expansion. See Table~\ref{examples} for some explicit examples.
    
    \item For a general regular singularity labeled by $f$, the polynomial  takes the following form:
    \begin{equation}\label{eq:general_case1}
       \boxed{ C^f_{m,n}(q,t) = \sum_\lambda f_\lambda(q,t) K_{\lambda f},}
    \end{equation}
    where  $f_\lambda(q,t)$  is from formula \ref{eq:parking} and $K_{\lambda f}$ denotes the Kostka number between two partitions $\lambda$ and $f$, see the Sage code in \cite{kostka} in computing it.
\end{enumerate}


We conjecture that \eqref{eq:general_case1} coincides with the bi-graded polynomial arising from either the mixed Hodge or perverse filtration. 
We also give a proposal to compute the bi-graded polynomial by using the geometry of affine Springer fiber following Hikita \cite{hikita2014affine}, which is used 
to verify the formula \ref{eq:general_case1}. The main formula for the computation is \ref{final}.

This paper is organized as follows: Section 2 presents the combinatorial definition of the bi-graded polynomials by using parking function; Section 3 gives a computation using the geometry of affine Springer fiber;
Section 4 compares our proposal with some known results from \cite{hausel2019arithmetic}; Section 5 provides concluding remarks.

\section{Rational parking function}
As discussed in introduction, the bigraded polynomial for the cohomology of Hitchin's system is determined combinatorial 
by the rational parking function \cite{armstrong2016rational}. We will review its definition and list many explicit computations. We also discuss 
examples for arbitrary regular singularity. 

\subsection{Full regular singularity: Rational parking function}
We consider the regular singularity $f$ of the form $[1,\ldots, 1]$, known as the full regular singularity. The bi-graded polynomial for the cohomology of the corresponding Hitchin's moduli space is given by the graded version of rational parking function \cite{armstrong2016rational}.

An $m/n$ rational parking function is defined as any $n$-tuple of non-negative integers $(f_1,\ldots, f_n)$ satisfying
\begin{equation*}
    \#\{i:~f_i<l\} \geq \frac{nl}{m}
\end{equation*}
for all $l \in [m]$. For example, $[0,1,4]$ is a $5/3$ parking function. Each rational parking function can be represented geometrically by a labeled $(m,n)$-Dyck path, which consists of:

\begin{itemize}
    \item An $(m,n)$-Dyck path: a lattice path from $(0,0)$ to $(m,n)$ staying strictly above the main diagonal of an $m\times n$ rectangle.
    \item A labeling of the north steps by $\{1,2,\ldots, n\}$ such that labels increase in each column when read from bottom to top. This defines a permutation $\sigma$ where $\sigma(i)$ is the label of the $i$th vertical step.
\end{itemize}

The width and height of an $(m,n)$-Dyck path are $m$ and $n$ respectively. There are exactly $m^{n-1}$ such labeled paths\footnote{The number of unlabeled $(m,n)$-Dyck paths is given by the rational Catalan number $\frac{(m+n-1)!}{m!n!}$.}. The symmetric group $S_n$ acts on $(m,n)$-parking functions by permuting labels  and rearranging the labels so that it is increasing within each column.

\begin{figure}[h]
\centering
\tikzset{every picture/.style={line width=0.75pt}} 

\begin{tikzpicture}[x=0.75pt,y=0.75pt,yscale=-1,xscale=1]

\draw   (201,151.22) -- (450,151.22) -- (450,300) -- (201,300) -- cycle ;
\draw    (250,150) -- (250,300) ;
\draw    (300,150) -- (300,300) ;
\draw    (350,150) -- (350,300) ;
\draw    (400,150) -- (400,300) ;
\draw    (200,300) -- (450,150) ;
\draw [line width=2.25]    (200,300) -- (200,250) ;
\draw [line width=2.25]    (200,250) -- (250,250) ;
\draw [line width=2.25]    (250,250) -- (250,200) ;
\draw [line width=2.25]    (250,150) -- (450,150) ;
\draw [line width=2.25]    (250,200) -- (250,150) ;
\draw    (200,200) -- (450,200) ;
\draw    (200,250) -- (450,250) ;
\draw  [color={rgb, 255:red, 155; green, 155; blue, 155 }  ,draw opacity=1 ][fill={rgb, 255:red, 80; green, 227; blue, 194 }  ,fill opacity=1 ] (300,150) -- (350,150) -- (350,200) -- (300,200) -- cycle ;
\draw  [color={rgb, 255:red, 155; green, 155; blue, 155 }  ,draw opacity=1 ][fill={rgb, 255:red, 80; green, 227; blue, 194 }  ,fill opacity=1 ] (250,150) -- (300,150) -- (300,200) -- (250,200) -- cycle ;

\draw (181,271.4) node [anchor=north west][inner sep=0.75pt]    {$2$};
\draw (233,214.4) node [anchor=north west][inner sep=0.75pt]    {$1$};
\draw (237,163.4) node [anchor=north west][inner sep=0.75pt]    {$3$};
\end{tikzpicture}
\caption{An example of a labeled $(m,n)$-Dyck path representing a $5/3$ parking function. The parking function from the unlabeled path is $[P_1,P_2, P_3]= [0,1,1]$ and the permutation from the label is $\sigma=213$, and the permuted one is $[P_{\sigma^{-1}(1)}, P_{\sigma^{-1}(2)}, P_{\sigma^{-1}(3)}]=[P_2,P_1,P_3]=[1,0,1]$.}
\label{fig:dyck_path}
\end{figure}
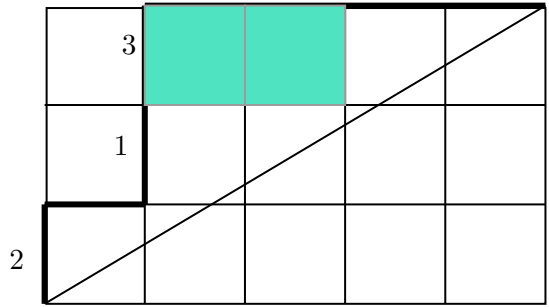

The bijection between $m/n$-parking functions and vertically labeled $m/n$-Dyck paths is constructed as follows: given a parking function $f=(f_1,f_2,\ldots, f_n)$, let $(P_1,P_2,\ldots, P_n)$ be its increasing rearrangement. The corresponding Dyck path has its $i$th vertical step at $x = P_i$, with the label given by the index of $P_i$ in $f$. 

The inverse map is straightforward: for a labeled Dyck path $(\sigma, P)$, the parking function is obtained by applying the permutation $\sigma$ to the sequence $(P_1, \ldots, P_n)$ of $x$-coordinates of the vertical steps: $[f_1,\ldots, f_n]=[P_{\sigma^{-1}(1)}, 
\ldots, P_{\sigma^{-1}(n)}]$.

Let $PF_{m,n}$ denote the set of all labeled $(m,n)$-parking functions. The Frobenius series is defined as:
\begin{equation*}
\boxed{PF_{m,n}(q,t) = \sum_{P\in PF(m,n)} t^{\text{area}(P)} q^{\text{dinv}(P)} F_{n,\text{lDes}(P)}(x)}
\end{equation*}

We now explain the various statistics appearing in this formula:

\begin{enumerate}
    \item \textbf{Area statistic}: The area of a parking function $P$ is defined as the number of complete lattice squares between the Dyck path and the main diagonal.
    For the example in figure. \ref{fig:dyck_path}, we have $area(P)=2$. 
    
    \item \textbf{Dinv statistic}: The dinv is the most difficult one to define, and it consists of three components:
    
    First, one need to define the rank of a lattice point $(x,y)$ at the start of a north step as:
    \begin{equation} \text{rank}((x,y)) = m y - n x 
    \label{rank}
    \end{equation}
    Notice that the rank only depends on the unlabelled Dyck path. Using the rank, we get a permutation $\sigma$ when we order the labels by \textbf{decreasing} rank.
    For example, the rank numbers for the path $P$ in figure. \ref{fig:dyck_path} is $[0,2,7]$, and so the permutation $\boxed{\sigma=312}$ as the label $3$ now has maximal rank $7$, etc.
    
    For the underlying Dyck path $P$, let $\lambda(P)$ denote the set of boxes \textbf{above} the path $P$. For any cell $x \in \lambda(P)$, define:
    \begin{itemize}
        \item $\text{leg}(x)$: number of cells in $\lambda(P)$ strictly south of $x$
        \item $\text{arm}(x)$: number of cells in $\lambda(P)$ strictly east of $x$
    \end{itemize}
    
    The \emph{path dinv} statistic is defined as:
    \begin{equation} \text{pdinv}(P) = \#\left\{ x \in \lambda(P) \;\middle|\; \frac{\text{arm}(x)}{\text{leg}(x)+1} < \frac{m}{n} < \frac{\text{arm}(x)+1}{\text{leg}(x)}
    \label{pdinv}
     \right\} \end{equation}
    This depends only on the Dyck path, not the labeling.
    
    The \emph{tdinv} counts the following pairs in the labeling of the Dyck path:
    \[ \text{tdinv}(P) = \#\left\{ (i,j) \;|\; i < j \text{ and } \text{rank}(i) < \text{rank}(j) < \text{rank}(i) + m \right\} \]
    where $i,j$ are labels in the path. 
    
    The \emph{maximal tdinv} is the maximal \emph{tdinv} with the same unlabeled path.
    
    The complete dinv  on a path is then:
    \[ \text{dinv}(P) = \text{pdinv}(P) + \text{tdinv}(P) - \text{maxtdinv}(P) \]
    
    \textbf{Example}:  Consider the example in figure. \ref{fig:dyck_path}, there is only one box above the path with $leg(x)=arm(x)=0$, so $pdinv=1$ by using formula \ref{pdinv} . The rank for the labelled path is $rank(1)=2, rank(2)=0, rank(3)=7$, and it is easy 
    to get $tdinv(P)=1$.  By looking at other labels with the same underlying Dyck path, we get $\text{maxtdinv}(P)=1$. We finally get $\text{dinv}(P)=1$.
    
    \item \textbf{Quasisymmetric function}: A formal power series in $x$ is quasisymmetric if it is invariant under index-order-preserving variable shifts. Gessel's fundamental basis is indexed by pairs $(n,S)$ where $S \subseteq \{1,2,\ldots, n-1\}$:
    \[ F_{n,S} = \sum_{\substack{1 \leq i_1 \leq \cdots \leq i_n \\ i_j < i_{j+1} \text{ when } j \in S}} x_{i_1} \cdots x_{i_n} \]

    Given a labeled Dyck path, one has a permutation $\sigma$ defined by the rank function. The set $S$ is defined as the inverse descent set from $S$.
    The inverse descent set $\text{IDes}(\sigma)$ of a permutation $\sigma$ is:
    \[ j \in \text{IDes}(\sigma) \iff j+1 \text{ appears left of } j \text{ in } \sigma \]
\end{enumerate}
For the permutation $\sigma=312$, the inverse descent set is just $2$. For more information about the quasi-symmetric polynomial, see appendix \ref{symmetric}.

The Frobenius series also admits a Schur expansion via the relation between Schur polynomials and quasisymmetric polynomials (see Appendix~\ref{symmetric}):
\begin{equation*}
\boxed{PF_{m,n}(q,t) = \sum_{\lambda \in P_n} f_\lambda(q,t) s_\lambda}
\end{equation*}

Some general remarks about the Frobenius series and the related Schur expansion:
\begin{enumerate}
\item  The number of terms in $PF_{m,n}$ is equal to $m^{n-1}$, which is equal to the dimension of the cohomology group of our Hitchin's moduli space. This is perhaps the first clue which leads me to 
conjecture the connection between the rational parking function and the bi-graded polynomial from Hitchin's moduli space.
\item  By setting $t=q=1$, $PF_{m,n}$ has the expansion 
\begin{equation*}
PF_{m,n}=\sum_\lambda f_\lambda (1,1) s_\lambda(x)
\end{equation*}
and $f_{1,1}$ is given as 
\begin{equation*}
f_\lambda(1,1)=\frac{1}{m}\prod_{x\in \lambda} \frac{m+c(x)}{h(x)}
\end{equation*}
where $h(x)$ is the Hook length of $x$ in Young Tableaux, and $c(x)$ is the content of $x$ (if x is in row $i$ and column $j$, and $c(x)=j-i$).
\item $PF_{m,n}(q,t)$ is conjectured to be symmetric in $q$ and $t$.
\end{enumerate}

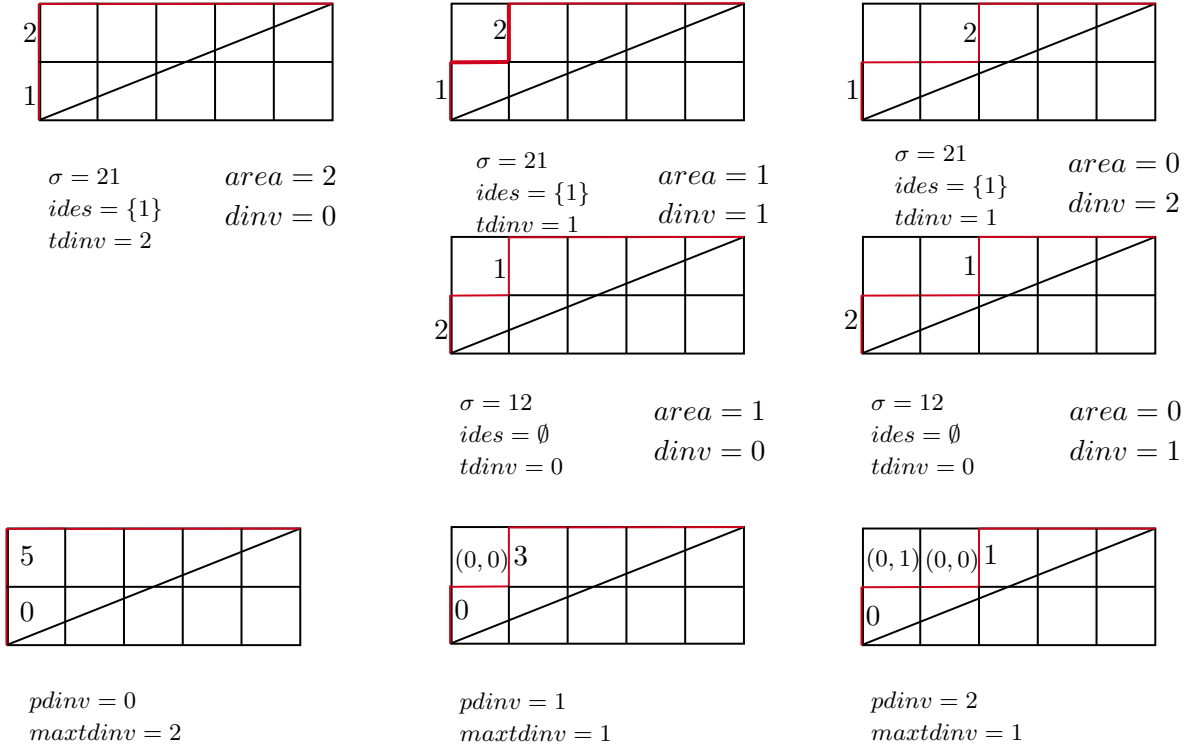
\begin{figure}

\begin{center}

\tikzset{every picture/.style={line width=0.75pt}} 

\begin{tikzpicture}[x=0.55pt,y=0.55pt,yscale=-1,xscale=1]

\draw   (121,80) -- (320,80) -- (320,160) -- (121,160) -- cycle ;
\draw    (120,160) -- (320,80) ;
\draw [color={rgb, 255:red, 208; green, 2; blue, 27 }  ,draw opacity=1 ]   (120,80.28) -- (160,80) ;
\draw [color={rgb, 255:red, 208; green, 2; blue, 27 }  ,draw opacity=1 ]   (120,120) -- (120,160) ;
\draw [color={rgb, 255:red, 208; green, 2; blue, 27 }  ,draw opacity=1 ]   (120,80) -- (120,120) ;
\draw [color={rgb, 255:red, 208; green, 2; blue, 27 }  ,draw opacity=1 ]   (160,80.28) -- (320,80) ;
\draw    (120,120) -- (320,120) ;
\draw    (280,160) -- (280,80) ;
\draw    (240,160) -- (240,80) ;
\draw    (200,160) -- (200,80) ;
\draw    (160,160) -- (160,120) ;
\draw   (401,80) -- (600,80) -- (600,160) -- (401,160) -- cycle ;
\draw    (400,160) -- (600,80) ;
\draw [color={rgb, 255:red, 208; green, 2; blue, 27 }  ,draw opacity=1 ][line width=1.5]    (400,120.28) -- (440,120) ;
\draw [color={rgb, 255:red, 208; green, 2; blue, 27 }  ,draw opacity=1 ]   (400,120) -- (400,160) ;
\draw [color={rgb, 255:red, 208; green, 2; blue, 27 }  ,draw opacity=1 ][line width=1.5]    (440,80) -- (440,120) ;
\draw [color={rgb, 255:red, 208; green, 2; blue, 27 }  ,draw opacity=1 ]   (440,80) -- (600,79.72) ;
\draw    (440,120) -- (600,120) ;
\draw    (560,160) -- (560,80) ;
\draw    (520,160) -- (520,80) ;
\draw    (480,160) -- (480,80) ;
\draw    (440,160) -- (440,120) ;
\draw    (160,120) -- (160,80) ;
\draw   (681,80) -- (880,80) -- (880,160) -- (681,160) -- cycle ;
\draw    (680,160) -- (880,80) ;
\draw [color={rgb, 255:red, 208; green, 2; blue, 27 }  ,draw opacity=1 ]   (680,120.28) -- (760,120) ;
\draw [color={rgb, 255:red, 208; green, 2; blue, 27 }  ,draw opacity=1 ]   (680,120) -- (680,160) ;
\draw [color={rgb, 255:red, 208; green, 2; blue, 27 }  ,draw opacity=1 ]   (760,80) -- (760,120) ;
\draw [color={rgb, 255:red, 208; green, 2; blue, 27 }  ,draw opacity=1 ]   (760,80) -- (880,80) ;
\draw    (760,120) -- (880,120) ;
\draw    (840,160) -- (840,80) ;
\draw    (800,160) -- (800,80) ;
\draw    (760,160) -- (760,120) ;
\draw    (720,160) -- (720,120) ;
\draw    (720,120) -- (720,80) ;
\draw   (401,240) -- (600,240) -- (600,320) -- (401,320) -- cycle ;
\draw    (400,320) -- (600,240) ;
\draw [color={rgb, 255:red, 208; green, 2; blue, 27 }  ,draw opacity=1 ]   (400,280.28) -- (440,280) ;
\draw [color={rgb, 255:red, 208; green, 2; blue, 27 }  ,draw opacity=1 ]   (400,280) -- (400,320) ;
\draw [color={rgb, 255:red, 208; green, 2; blue, 27 }  ,draw opacity=1 ]   (440,240) -- (440,280) ;
\draw [color={rgb, 255:red, 208; green, 2; blue, 27 }  ,draw opacity=1 ]   (440,240.28) -- (600,240) ;
\draw    (440,280) -- (600,280) ;
\draw    (560,320) -- (560,240) ;
\draw    (520,320) -- (520,240) ;
\draw    (480,320) -- (480,240) ;
\draw    (440,320) -- (440,280) ;
\draw   (681,240) -- (880,240) -- (880,320) -- (681,320) -- cycle ;
\draw    (680,320) -- (880,240) ;
\draw [color={rgb, 255:red, 208; green, 2; blue, 27 }  ,draw opacity=1 ]   (680,280.28) -- (760,280) ;
\draw [color={rgb, 255:red, 208; green, 2; blue, 27 }  ,draw opacity=1 ]   (680,280) -- (680,320) ;
\draw [color={rgb, 255:red, 208; green, 2; blue, 27 }  ,draw opacity=1 ]   (760,240) -- (760,280) ;
\draw [color={rgb, 255:red, 208; green, 2; blue, 27 }  ,draw opacity=1 ]   (760,240) -- (880,240) ;
\draw    (760,280) -- (880,280) ;
\draw    (840,320) -- (840,240) ;
\draw    (800,320) -- (800,240) ;
\draw    (760,320) -- (760,280) ;
\draw    (720,320) -- (720,280) ;
\draw    (720,280) -- (720,240) ;
\draw   (99,440) -- (298,440) -- (298,520) -- (99,520) -- cycle ;
\draw    (98,520) -- (298,440) ;
\draw [color={rgb, 255:red, 208; green, 2; blue, 27 }  ,draw opacity=1 ]   (98,440.28) -- (138,440) ;
\draw [color={rgb, 255:red, 208; green, 2; blue, 27 }  ,draw opacity=1 ]   (98,480) -- (98,520) ;
\draw [color={rgb, 255:red, 208; green, 2; blue, 27 }  ,draw opacity=1 ]   (98,440) -- (98,480) ;
\draw [color={rgb, 255:red, 208; green, 2; blue, 27 }  ,draw opacity=1 ]   (138,440.28) -- (298,440) ;
\draw    (98,480) -- (298,480) ;
\draw    (258,520) -- (258,440) ;
\draw    (218,520) -- (218,440) ;
\draw    (178,520) -- (178,440) ;
\draw    (138,520) -- (138,480) ;
\draw    (138,480) -- (138,440) ;
\draw   (401,439) -- (600,439) -- (600,519) -- (401,519) -- cycle ;
\draw    (400,519) -- (600,439) ;
\draw [color={rgb, 255:red, 208; green, 2; blue, 27 }  ,draw opacity=1 ]   (400,480) -- (440,479.72) ;
\draw [color={rgb, 255:red, 208; green, 2; blue, 27 }  ,draw opacity=1 ]   (400,479) -- (400,519) ;
\draw [color={rgb, 255:red, 208; green, 2; blue, 27 }  ,draw opacity=1 ]   (440,439) -- (440,479) ;
\draw [color={rgb, 255:red, 208; green, 2; blue, 27 }  ,draw opacity=1 ]   (440,439.28) -- (600,439) ;
\draw    (440,480) -- (600,480) ;
\draw    (560,519) -- (560,439) ;
\draw    (520,519) -- (520,439) ;
\draw    (480,519) -- (480,439) ;
\draw    (440,519) -- (440,479) ;
\draw   (681,440) -- (880,440) -- (880,520) -- (681,520) -- cycle ;
\draw    (680,520) -- (880,440) ;
\draw [color={rgb, 255:red, 208; green, 2; blue, 27 }  ,draw opacity=1 ]   (680,480.28) -- (760,480) ;
\draw [color={rgb, 255:red, 208; green, 2; blue, 27 }  ,draw opacity=1 ]   (680,480) -- (680,520) ;
\draw [color={rgb, 255:red, 208; green, 2; blue, 27 }  ,draw opacity=1 ]   (760,440) -- (760,480) ;
\draw [color={rgb, 255:red, 208; green, 2; blue, 27 }  ,draw opacity=1 ]   (760,440) -- (880,440) ;
\draw    (760,480) -- (880,480) ;
\draw    (840,520) -- (840,440) ;
\draw    (800,520) -- (800,440) ;
\draw    (760,520) -- (760,480) ;
\draw    (720,520) -- (720,480) ;
\draw    (720,480) -- (720,440) ;

\draw (107,134.4) node [anchor=north west][inner sep=0.75pt]    {$1$};
\draw (107,91.4) node [anchor=north west][inner sep=0.75pt]    {$2$};
\draw (387,130.4) node [anchor=north west][inner sep=0.75pt]    {$1$};
\draw (427,89.4) node [anchor=north west][inner sep=0.75pt]    {$2$};
\draw (427,253.4) node [anchor=north west][inner sep=0.75pt]    {$1$};
\draw (387,295.4) node [anchor=north west][inner sep=0.75pt]    {$2$};
\draw (747,91.4) node [anchor=north west][inner sep=0.75pt]    {$2$};
\draw (667,130.4) node [anchor=north west][inner sep=0.75pt]    {$1$};
\draw (747,251.4) node [anchor=north west][inner sep=0.75pt]    {$1$};
\draw (666,288.4) node [anchor=north west][inner sep=0.75pt]    {$2$};
\draw (241,183.4) node [anchor=north west][inner sep=0.75pt]    {$ \begin{array}{l}
area=2\\
dinv=0
\end{array}$};
\draw (536,182.4) node [anchor=north west][inner sep=0.75pt]    {$ \begin{array}{l}
area=1\\
dinv=1
\end{array}$};
\draw (815,174.4) node [anchor=north west][inner sep=0.75pt]    {$ \begin{array}{l}
area=0\\
dinv=2
\end{array}$};
\draw (533,344.4) node [anchor=north west][inner sep=0.75pt]    {$ \begin{array}{l}
area=1\\
dinv=0
\end{array}$};
\draw (816,344.4) node [anchor=north west][inner sep=0.75pt]    {$ \begin{array}{l}
area=0\\
dinv=1
\end{array}$};
\draw (105,488.4) node [anchor=north west][inner sep=0.75pt]    {$0$};
\draw (105,450.4) node [anchor=north west][inner sep=0.75pt]    {$5$};
\draw (401,487.4) node [anchor=north west][inner sep=0.75pt]    {$0$};
\draw (441,450.4) node [anchor=north west][inner sep=0.75pt]    {$3$};
\draw (761,449.4) node [anchor=north west][inner sep=0.75pt]    {$1$};
\draw (681,491.4) node [anchor=north west][inner sep=0.75pt]    {$0$};
\draw (412,175.4) node [anchor=north west][inner sep=0.75pt]    [font=\footnotesize] {$ \begin{array}{l}
\sigma =21\\
ides=\{1\}\\
tdinv=1
\end{array}$};
\draw (401,341.4) node [anchor=north west][inner sep=0.75pt]    [font=\footnotesize] {$ \begin{array}{l}
\sigma =12\\
ides=\emptyset \\
tdinv=0
\end{array}$};
\draw (108,545.4) node [anchor=north west][inner sep=0.75pt]   [font=\footnotesize]  {$ \begin{array}{l}
pdinv=0\\
maxtdinv=2
\end{array}$};
\draw (401,546.4) node [anchor=north west][inner sep=0.75pt]    [font=\footnotesize] {$ \begin{array}{l}
pdinv=1\\
maxtdinv=1
\end{array}$};
\draw (681,545.4) node [anchor=north west][inner sep=0.75pt]    [font=\footnotesize] {$ \begin{array}{l}
pdinv=2\\
maxtdinv=1
\end{array}$};
\draw (401,452.4) node [anchor=north west][inner sep=0.75pt]  [font=\footnotesize]  {$( 0,0)$};
\draw (681,451.4) node [anchor=north west][inner sep=0.75pt]  [font=\footnotesize]  {$( 0,1)$};
\draw (721,452.4) node [anchor=north west][inner sep=0.75pt]  [font=\footnotesize]  {$( 0,0)$};
\draw (121,186.4) node [anchor=north west][inner sep=0.75pt]   [font=\footnotesize]  {$ \begin{array}{l}
\sigma =21\\
ides=\{1\}\\
tdinv=2
\end{array}$};
\draw (681,341.4) node [anchor=north west][inner sep=0.75pt]   [font=\footnotesize]  {$ \begin{array}{l}
\sigma =12\\
ides=\emptyset \\
tdinv=0
\end{array}$};
\draw (697,170.4) node [anchor=north west][inner sep=0.75pt]   [font=\footnotesize]  {$ \begin{array}{l}
\sigma =21\\
ides=\{1\}\\
tdinv=1
\end{array}$};

\end{tikzpicture}

\caption{The combinatorial data for rational parking function with $m=5, n=2$.}
\label{rank2}
\end{center}

\end{figure}

\textbf{Example}: Let's take $m=5, n=2$, the set of data for computing the graded rational parking function are listed in figure. \ref{rank2}, and so
\begin{equation*}
PF_{5,2}=(q^2+qt+t^2)s_2+(q +t)s_{1,1}
\end{equation*}
Here $s_2, s_{1,1}$ are the Schur polynomial in two variables.

Some explicit examples are computed in \cite{leven2014two,qiu2020schur}, and we list them in table. \ref{examples}. 
\begin{table}[htp]
\begin{center}
\begin{tabular}{|c|c|} \hline
$C_{2k+1,2}(q,t)= [k]_{q,t}s_2 + [k+1]_{q,t}s_{11}$ \\ \hline
 $C_{2,2k+1}(q,t)= \sum_{r=0}^k[k + 1 - r]_{q,t} \, s_{2^r1^{2k+1-2r}} $\\ \hline
$C_{3k+1,3}(q,t)=(\sum_{i=0}^{k-1} s_{k+2i-1, k-i-1}(q,t))s_{(3)}+(\sum_{i=0}^{k-1}[s_{k+2i,k-i-1}(q,t)+$  \\ 
~~~~~~~~$s_{k+2i+1,k-i-1}(q,t)]s_{(21)}+(\sum_{i=0}^ks_{k+2i, k-i}(q,t)) s_{(111)}$  \nonumber\\ \hline
$C_{3k+2,3}(q,t)=(\sum_{i=0}^{k-1} s_{k+2i, k-i-1}(q,t))s_{(3)}+(\sum_{i=-1}^{k-1}[s_{k+2i+1,k-i-1}(q,t)+ $ \\ 
~~~~~~~~$s_{k+2i+2,k-i-1}(q,t)]s_{(21)}+(\sum_{i=0}^ks_{k+2i+1, k-i}(q,t)) s_{(111)}$ \\ \hline
\end{tabular}
\end{center}
\caption{The two parameter $q,t$ polynomial $s_{(a,b)}(q, t) = (qt)^b [a - b + 1]_{q,t}$, and $[k]_{q,t}=\frac{q^k-t^k}{q-t}=q^{k-1}+q^{k-2}t+\ldots+qt^{k-2}+t^{k-1}$.}
\label{examples}
\end{table}%

\subsection{General Regular Singularity}

Consider the polynomial \( C_{m,n}(q,t) = \sum_\lambda f_\lambda s_\lambda \) associated with the maximal regular singularity. For general regular singularity labeled by a partition $f \in P_n$, the bi-graded polynomial generalizes to:
\begin{equation}
    C^f_{m,n}(q,t) = \sum_\lambda f_\lambda(q,t) K_{\lambda f},
    \label{eq:general_case}
\end{equation}
where \( K_{\lambda f} \) denotes the Kostka number. When \( f = [1, \ldots, 1] \), \( K_{\lambda f} \) coincides with the dimension of the irreducible representation corresponding to the Young tableau \( \lambda \), which gives back the answer
for the maximal regular singularity. 

The Kostka number is closely tied to the parameterization of the coset \( S_n/S_f \), where \( S_n \) is the symmetric group and \( S_f \) is the parabolic subgroup associated with \( f \). This is particularly relevant because fixed points on the affine Springer fiber are also parameterized by the coset \( S_n/S_f \) \cite{oblomkov2016geometric}, motivating our conjecture.

The following remarks clarify key aspects of our results:
\begin{enumerate}
    \item When \( f = [n] \), the only non-zero Kostka number is \( K_{[n],[n]} = 1 \), yielding
    \[
        C^{[n]}(q,t) = f_{[n]}(q,t).
    \]
    
    
    \item For \( f = [n-1, 1] \), the non-zero Kostka numbers are \( K_{[n],[n-1,1]} = 1 \) and \( K_{[n-1,1],[n-1,1]} = 1 \), leading to
    \[
        C^{[n-1,1]}(q,t) = f_{[n-1,1]}(q,t) + f_{[n]}(q,t).
    \]
    
    \item When \( m < n \), the Schur expansion is truncated, and the final term \( \lambda_{\text{max}} \) in the expansion corresponds to the orbit of the associated variety \cite{MR3456698}. Physically, this orbit’s closure defines the Higgs branch \cite{Xie:2019vzr}. For \( f = \lambda_{max} \), the Hitchin moduli space reduces to a point, so the coefficient of \( S_{\lambda_{max}}\) must be \( 1 \). The polynomial takes the form:
    \[
        C_{m,n}(q,t) = \ldots + [1]s_{[m,m,\ldots,m,s]} \quad \text{for } m < n.
    \]
\end{enumerate}

\textbf{Example 1}: Consider $n=2,~m=2k+1$, and there are two types of regular punctures with partition $f=[1,1]$ and $f=[2]$. 

For $f=[1,1]$, the associated bigraded polynomial is 
\begin{equation*}
C_{2k+1, 2}(q,t)=[k]_{q,t}s_{[2]}+[k+1]_{q,t} s_{[1,1]}
\end{equation*}
and so the conjectural polynomial on the cohomology is 
\begin{align*}
&H(q,t)=q^k t^{2k} C_{2k+1, 2}(q, q^{-1} t^{-2})=q^k t^{2k}([k]_{q,q^{-1}t^{-2}}s_{[2]}+[k+1]_{q,q^{-1}t^{-2}} s_{[1,1]})= \nonumber\\
&q^k t^{2k}[\sum_{i=0}^{k-1} q^i (q^{-1} t^{-2})^{k-1-i} s_{[2]}+\sum_{i=0}^{k} q^i (q^{-1} t^{-2})^{k-i}  s_{[1,1]})] = \nonumber\\
& \boxed{ \sum_{i=0}^{k-1} q^{2i+1}t^{2i+2}  s_{[2]}+\sum_{i=0}^{k} q^{2i}t^{2i}  s_{[1,1]}}
\end{align*}
by using the result the dimension of Hitchin moduli space is $d=k$.

For $f=[2]$, the associated bi-grade polynomial is simply $C_{2k+1,2}^{[2]}(q,t)=[k]_{q,t}$, and so the conjectural polynomial on the cohomology is 
\begin{align*}
&H(q,t)=q^{k-1} t^{2(k-1)}([k]_{q,q^{-1}t^{-2}})=q^{k-1} t^{2(k-1)}\sum_{i=0}^{k-1} q^i (q^{-1} t^{-2})^{k-1-i} \nonumber\\
&=\boxed{\sum_{i=0}^{k-1}q^{2i}t^{2i}}
\end{align*}
by using the result that the dimension of the Hitchin moduli space is $d=k-1$.

\textbf{Example 2}: Consider \( n = 5 \), \( m = 3 \), with the graded rational parking function:
\[
    C_{3,5}(q,t) = 
    \left[ \begin{array}{ccccc} 
        \cdot & \cdot & \cdot & \cdot & 1 \\  
        \cdot & \cdot & 1 & 1 & \cdot \\
        \cdot & 1 & 1 & \cdot & \cdot \\
        \cdot & 1 & \cdot & \cdot & \cdot \\
        1 & \cdot & \cdot & \cdot & \cdot  
    \end{array} \right]s_{[1,1,1,1,1]} +
    \left[ \begin{array}{cccc} 
        \cdot & \cdot & 1 & 1 \\
        \cdot & 2 & 1 & \cdot  \\
        1 & 1 & \cdot & \cdot  \\
        1 & \cdot & \cdot & \cdot   
    \end{array} \right]s_{[2,1,1,1]} +
    \left[ \begin{array}{ccc} 
        \cdot & 1 & 1 \\
        1 & 1 & \cdot  \\
        1 & \cdot & \cdot 
    \end{array} \right]s_{[2,2,1]} +
    \left[ \begin{array}{cc} 
        \cdot & 1 \\
        1 & \cdot  
    \end{array} \right]s_{[3,1,1]} +
    [1]s_{[3,2]}.
\]
The matrix entries encode \( q,t \)-monomials: for an \( (i,j) \) entry $a_{ij}$, the term is \( a_{ij} q^{i-1} t^{j-1} \). The Kostka numbers for \( S_5 \) group are listed in Table \ref{s5kos}, and we get (using formula \ref{eq:general_case}):
\[
    C_{3,5}^{[2,1,1,1]} = 
    \left[ \begin{array}{cccc} 
        \cdot & \cdot & 1 & 1 \\
        \cdot & 2 & 1 & \cdot  \\
        1 & 1 & \cdot & \cdot  \\
        1 & \cdot & \cdot & \cdot   
    \end{array} \right]\times 1
    + \left[ \begin{array}{ccc} 
        \cdot & 1 & 1 \\
        1 & 1 & \cdot  \\
        1 & \cdot & \cdot 
    \end{array} \right] \times 2
    + \left[ \begin{array}{cc} 
        \cdot & 1 \\
        1 & \cdot  
    \end{array} \right] \times 3 + [1] \times 3,
\]
\[
    C_{3,5}^{[2,2,1]} = 
    \left[ \begin{array}{ccc} 
        \cdot & 1 & 1 \\
        1 & 1 & \cdot  \\
        1 & \cdot & \cdot 
    \end{array} \right]\times 1
    + \left[ \begin{array}{cc} 
        \cdot & 1 \\
        1 & \cdot  
    \end{array} \right] \times 1 + [1] \times 2,
\]
\[
    C_{3,5}^{[3,1,1]} = 
    \left[ \begin{array}{cc} 
        \cdot & 1 \\
        1 & \cdot  
    \end{array} \right] \times 1 + [1] \times 1,
\]
\[
    C_{3,5}^{[3,2]} = [1] \times 1.
\]

Some observations:
a) The number of terms in each polynomial matches those listed in Table \ref{tab:dimensions} (for \( m = 3 \), \( n = 5 \)). For \( f = [2,1,1,1] \), the exponents \( e = (1,0,0,0) \); for \( f = [2,2,1] \), \( e = (1,1,0,0) \); and for \( f = [3,2] \), \( e = (1,2,1,0) \). 

b) For \( f = [3,1,1] \), the Coulomb branch spectrum \cite{Xie:2012hs} (e.g., a single operator with scaling dimension \( \frac{4}{3} \)) implies the IR theory is equivalent to the \( sl_2 \) Hitchin system with \( m = 3 \), \( n = 2 \), and \( f = [1,1] \). Thus, the bi-graded polynomial should be \( C_{3,2}(q,t) = [2]_{q,t} + [1]_{q,t} = q + t + 1 \), matching the result in the preceding equations (see Table \ref{examples} and set Schur polynomial fugacity to 1).

\begin{table}[htp]
\caption{Kostka numbers for the \( S_5 \) group.}
\centering
\begin{tabular}{|c|c|c|c|c|c|c|c|} \hline
~ & \( [5] \) & \( [4,1] \) & \( [3,2] \) & \( [3,1,1] \) & \( [2,2,1] \) & \( [2,1,1,1] \) & \( [1,1,1,1,1] \) \\ \hline
\( [5] \) & 1 & 1 & 1 & 1 & 1 & 1 & 1 \\ \hline
\( [4,1] \) & 0 & 1 & 1 & 2 & 2 & 3 & 4 \\ \hline
\( [3,2] \) & 0 & 0 & 1 & 1 & 2 & 3 & 5 \\ \hline
\( [3,1,1] \) & 0 & 0 & 0 & 1 & 1 & 3 & 6 \\ \hline
\( [2,2,1] \) & 0 & 0 & 0 & 0 & 1 & 2 & 5 \\ \hline
\( [2,1,1,1] \) & 0 & 0 & 0 & 0 & 0 & 1 & 4 \\ \hline
\( [1,1,1,1,1] \) & 0 & 0 & 0 & 0 & 0 & 0 & 1 \\ \hline
\end{tabular}
\label{s5kos}
\end{table}

\section{Affine Springer fiber}

We have defined our bigraded polynomial using Dyck paths. In this section, we will reformulate the same result using affine Springer fibers and the bigraded structure on cohomology proposed by Hikita \cite{hikita2014affine}.

Let $F = \mathbb{C}((\epsilon))$ be the field of Laurent power series in $\epsilon$, and $\mathcal{O} = \mathbb{C}[[\epsilon]]$ be the ring of formal power series in $\epsilon$. Given a Lie group $G$, we have the corresponding affine Lie group $G(F)$ and affine Lie algebra $\mathfrak{g}(F)$. The affine Grassmannian is defined as $G(F)/G(\mathcal{O})$. For an element $\gamma \in \mathfrak{g}(F)$, the affine Springer fiber is defined as
\begin{equation*}
    X_\gamma = \{ g \in G(F)/G(\mathcal{O}) \mid \mathrm{Ad}(g)^{-1}(\gamma) \in \mathfrak{g}(\mathcal{O}) \}.
\end{equation*}

Let's use $\Delta$ to denote the set of the roots of the finite Lie algebra $\mathfrak{g}$, and $\Delta_{+}$ the set of positive roots.
For a set of positive roots $\Delta_P \subset \Delta_+$, we define a parabolic subalgebra by
\begin{equation*}
    \mathfrak{g}_P = \left( \mathfrak{h} \oplus_{\alpha \in \Delta_P} (\mathfrak{g}_\alpha \oplus \mathfrak{g}_{-\alpha}) \right) \oplus \mathfrak{n}_P,
\end{equation*}
where the nilpotent part is $\mathfrak{n}_P= \oplus_{\alpha \notin \Delta_P} \mathfrak{g}_\alpha$.

In the $\mathfrak{sl}_n$ case, parabolic subalgebras are classified by partitions $f=[n_1, n_2, \ldots, n_s]$. The set of roots $\Delta_P$ is generated by
\begin{equation*}
    \Delta_P = \{\alpha_1, \ldots, \alpha_{n_1-1}\} \cup \{\alpha_{n_1+1}, \ldots, \alpha_{n_1+n_2-1}\} \cup \cdots.
\end{equation*}
In our notation:
\begin{itemize}
    \item The partition $[n]$ (with $\Delta_P = \Delta_+$) corresponds to the full Lie algebra.
    \item The partition $[1, \ldots, 1]$ (with $\Delta_P = \emptyset$) corresponds to Borel subalgebra.
\end{itemize}
Interestingly, these partitions coincide with the partitions $f$ appearing in our Hitchin system. There is an associated Weyl group $W_P$ generated by the simple reflections corresponding to roots in $\Delta_P$.

We define an element $\rho_P$ acting on simple roots by
\begin{equation*}
    \rho_P(\alpha) = 
    \begin{cases} 
        0 & \text{if } \alpha \in \Delta_P, \\
        1 & \text{if } \alpha \notin \Delta_P.
    \end{cases}
\end{equation*}

For a fixed parabolic subalgebra $\mathfrak{g}_P$, we can define the associated affine flag manifold. The corresponding affine Lie algebra $\widetilde{\mathfrak{g}}_P$ is generated by the roots $\alpha, \alpha \in \mathfrak{g}_P$ and $\alpha+k\delta, k>0,~\alpha \in \Delta$. We now define the generalized Springer fiber as
\begin{equation*}
    X_{P,\gamma} = \{ g \in G(F)/G_P \mid \mathrm{Ad}(g)^{-1}(\gamma) \in \widetilde{\mathfrak{g}}_P \}.
\end{equation*}
For our purposes, we introduce the affine Spaltenstein fiber:
\begin{equation*}
    B_{P,\gamma} = \{ g \in G(F)/G_P \mid \mathrm{Ad}(g)^{-1}(\gamma) \in \widetilde{\mathfrak{n}}_P \},
\end{equation*}
where $\widetilde{\mathfrak{n}}_P$ denotes the nilpotent part. 

We are interested in the so-called elliptic regular semi-simple element $\gamma$ \cite{oblomkov2016geometric, shan2023mirror}, and such elements are denoted by the slope ${\nu\over h}$ with $(\nu, h)=1$.
Let's take $\nu = ah + b$ with $\gcd(a, h) = 1,~~0<b<h$, we define the set of height $\nu$ affine roots \footnote{The height for an affine root $\alpha+k\delta$ is $kh+height(\alpha)$, and the height of a root is given by $height(\alpha)=\rho^\vee(\alpha)$.}:
\begin{equation}
    \widetilde{\Phi}_\nu = \{ \alpha + a\delta \mid \alpha \in \Phi_b \} \cup \{ \alpha + (a+1)\delta \mid \alpha \in \Phi_{h-b} \}.
    \label{nuset}
\end{equation}
with $\Phi_b$ ($\Phi_{h-b}$) the set of height $b$ (height $h-b$) roots. This set of roots are important in computing the cohomology of the affine Springer fiber.

\subsection{Two gradings on cohomology groups}
\subsubsection{Dimension of the Cell}
The cohomology groups of generalized Springer fibers are computed in \cite{goresky2006purity} using the affine Bruhat decomposition of the affine Grassmannian. The affine Spaltenstein fiber admits an affine paving, where the non-empty cells are characterized by the condition:
\begin{equation}
\boxed{\tilde{w} \in \widetilde{W}/W_P,~~ \tilde{w}(\Phi_\nu) \in \widetilde{\mathfrak{n}}_P}
\label{cell}
\end{equation}
Here, $\widetilde{W}$ denotes the affine Weyl group, $W_P$ is the parabolic Weyl group of $P$, and $\widetilde{\mathfrak{n}}_P$ is given by:
\begin{equation*}
\widetilde{\mathfrak{n}}_P = \{\alpha \mid \alpha \in \Delta_+ \setminus \Delta_P\} \cup \{\alpha + k\delta \mid k > 0\}.
\end{equation*}

The dimension of each cell is determined by counting positive affine roots $(\alpha, k)$ with height less than $\nu$, subject to additional inversion conditions \cite{goresky2006purity} 
stated below.

For a generic regular singularity labeled by a nilpotent orbit $f$ (corresponding to a partition of $n$ and gives rise to a parabolic subgroup $P$), we have ($\tilde{w}$ obeys \ref{cell}):
\begin{itemize}    
    \item The cell dimension $C_{\tilde{w}}$ is computed by counting affine roots satisfying the following conditions for an affine Weyl group element $\tilde{w}=st_q$:
    \begin{equation}
   dim(C_{\tilde{w}})=\#\boxed{ \{ (\alpha, k)|0 < (\rho, \alpha) + h k < \nu, \quad \& (-q, \alpha) + k + c(\rho_P, s(\alpha)) < 0\}}
   \label{celldim}
    \end{equation}
    where $0 < c \ll 1$. The first condition counts the positive affine roots with height less than $\nu$.
     The second equality is simplified to:
    \begin{equation}
    \begin{cases}
    \text{(a)} & (\rho_P, s(\alpha)) \geq 0 \text{ and } k + (-q, \alpha) < 0, \\
    \text{(b)} & (\rho_P, s(\alpha)) < 0 \text{ and } k + (-q, \alpha) \leq 0.
    \end{cases}
    \end{equation}
    Equivalently, these conditions can be expressed as:
    \begin{equation}
    \begin{cases}
    \text{(a)} & s(\alpha) \in \mathfrak{g}_P \text{ and } k + (-q, \alpha) < 0, \\
    \text{(b)} & s(\alpha) \in \Delta \setminus \mathfrak{g}_P \text{ and } k + (-q, \alpha) \leq 0.
    \end{cases}
    \label{dimension}
    \end{equation}
\end{itemize}

For the affine Spaltenstein variety, the cell dimension $C_{\tilde{w}}$ is given by the number of affine roots satisfy the condition \ref{celldim}, subtract the number of affine roots satisfying the following condition
\begin{equation}
\{k + (-q, \alpha)=0 ~\&~ s(\alpha) \in \Delta_P ~ \&~ height (\tilde{\alpha})\geq \nu \}
   \label{celldim1}
\end{equation}

\paragraph{Grading Formula}
The $t$-grading corresponds to the codimension of the cell, with the graded formula given by:
\begin{equation*}
\sum_{\tilde{w}} t^{d - \dim(C_{\tilde{w}})}.
\end{equation*}
where $d$ is the hyperkhaler dimension of the moduli space:
\begin{equation}
d=\frac{(\nu-1)(n-1)}{2}-dim{\cal O}_f
\label{modulidim}
\end{equation}

\subsubsection{The Second Grading}
The second grading arises from the following construction \cite{goresky2006purity}. Let's first take the trivial regular singularity so $f=[1,\ldots, 1]$. A $\nu$-stable affine Weyl group element is defined by the condition:
\begin{equation*}
\boxed{\tilde{w}(\tilde{\Phi}_\nu) \subset \tilde{\Phi}^+}
\end{equation*}
See \ref{nuset} for the definition of $\tilde{\Phi}_\nu$, and  $\tilde{\Phi}^+$ is the set of positive affine roots.
We denote such elements by $W^\nu$, and their inverses by ${}^\nu W$.

\paragraph{Alcove Geometry}
Let $V$ be the vector space generated by the roots.
The fundamental alcove is defined as:
\begin{equation*}
A_0 = \{ x \in V \mid \langle x, \alpha \rangle > 0 \text{ for all } \alpha \in \Delta \text{ and } \langle x, \theta \rangle < 1 \}
\end{equation*}
where $\theta$ is the highest root. The Sommers region consists of alcoves $\tilde{w}\Delta_0$ by using $\tilde{w} \in {}^\nu W$, see \cite{sommers1997family}. 

An important related region is the $\nu$-dilated fundamental alcove $\nu \Delta_0$. The connection between these regions is established through the set of roots:
\begin{equation*}
\tilde{\Delta}_{\nu} = \{-\theta + \nu \delta, \alpha_1, \ldots, \alpha_r\}.
\end{equation*}
There exists a unique affine Weyl group element $\tilde{w}_\nu$ such that:
\begin{equation*}
\tilde{w}_\nu (\tilde{\Phi}_\nu) = \tilde{\Delta}_\nu.
\end{equation*}

We define special Weyl group elements $\tilde{w}_d$ satisfying:
\begin{equation*}
\tilde{w}_d(\Delta_\nu) \subset \tilde{\Phi}^+.
\end{equation*}
The action of the inverse of these affine Weyl group elements on the fundamental alcove remains in the $\nu$-dilated region.
These $\tilde{w}_d $ and $\tilde{w}_s \in W^\nu$ elements are related by:
\begin{equation*}
\tilde{w}_d = \tilde{w}_s \tilde{w}^{-1}_\nu.
\end{equation*}

There exists a crucial map between ${}^\nu W$ and the coset $Q^\vee/\nu Q^\vee$ \cite{thiel2016anderson}:
\begin{equation}
{\cal A}: ~ W^\nu \to Q^\vee/\nu Q^\vee:~~\tilde{w}_s\to \boxed{\tilde{w}_s\tilde{w}^{-1}_\nu \cdot(0)}
\label{anderson}
\end{equation}
where $0$ is the origin of the roots space, and the action of an affine Weyl group element $\tilde{w}=t_q s$ acts on a root vector $\alpha$ as 
$t_q s\cdot (\alpha)=q+s(\alpha)$.

The Anderson map gives following useful geometric characterization of the set $W^\nu$:
\begin{enumerate}
    \item The Weyl group orbit representatives in the set $W^\nu$ correspond to lattice points in the closure of $\nu\Delta_0$.
    
    \item The stabilizer subgroup for a coroot lattice point in $\overline{\nu \Delta_0}$ is generated by hyperplanes containing that point. The orbit counting \cite{haiman1994conjectures, thiel2016anderson} yields:
    \begin{itemize}
        \item[(a)] The number of regular orbits:
        \begin{equation*}
        \frac{1}{|W|} \prod_{i=1}^l (\nu - e_i)
        \end{equation*}
        
        \item[(b)] The total number of orbits:
        \begin{equation*}
        \frac{1}{|W|} \prod_{i=1}^l (\nu + e_i)
        \end{equation*}
    \end{itemize}
    Here $e_i$ are the exponents of the Lie algebra and $|W|$ is the order of the Weyl group.
\end{enumerate}

\paragraph{Parking Function Correspondence}
The Anderson map can be used to define a map from the set $W^\nu$ to the set of parking function defined in last section.
Let's explain this map in some detail. First, 
for $\mathfrak{sl}_n$, the root space consists of vectors $(x_1,\ldots,x_n)$ with $\sum x_i = 0$. The simple roots are $\alpha_i = x_i - x_{i+1}$ for $i=1,\ldots,n-1$. 

Secondly, there exists a map $\zeta: Q^\vee/\nu Q^\vee \to \mathbb{Z}_\nu^n/\lambda(1,\ldots,1)$. By selecting the minimal element in each $\lambda(1,\ldots,1)$ orbit - a tuple $[a_1,\ldots,a_n]$ with $a_i \geq 0$ - we obtain a mapping from ${}^\nu W$ to parking functions $[a_1,\ldots,a_n]$.

\begin{example}
Consider $\mathfrak{sl}_3$ with $\nu=4$. Take the coroot $\alpha_1 + \alpha_2 = x_1 - x_3$, with coordinates $[1,0,-1] \equiv [1,0,3]$ in $\mathbb{Z}_4^3=\mathbb{Z}_4\times \mathbb{Z}_4 \times \mathbb{Z}_4$. Acting by $[1,1,1]$, we obtain the orbit:
\begin{equation*}
[2,1,0],\ [3,2,1],\ [0,3,2],\ [1,0,3].
\end{equation*}
The minimal element $[2,1,0]$ corresponds to the parking function of the coroot $\alpha_1+\alpha_2$. 
\end{example}

\paragraph{The q-Statistic}
Now we use the parking function to define the $q$ statistics on an element in the set $W^\nu$.
Given an element $\tilde{w} \in W^\nu$ and its related parking function $[a_1,\ldots, a_n]$, We define the $q$-grading as:
\begin{equation}
\boxed{qstat(\tilde{w}) = d - \sum a_i}
\end{equation}
where $d$ is the dimension of Hitchin's moduli space with regular singularity $[1,\ldots,1]$, namely
\begin{equation*}
d=\frac{(n-1(\nu-1)}{2}
\end{equation*}

\textbf{Remark}:  For an affine Weyl group element $\tilde{w} = st_q$, the $q$-statistic depends only on the root vector $q$.

\paragraph{Bigraded Polynomial}
The complete bigraded polynomial for the general affine Spalstein fiber $B_{P,\gamma}$ is given by:
\begin{equation}
\boxed{C_{\nu,n}^f(q,t) = \sum_{\tilde{w}} q^{\frac{(n-1((\nu-1)}{2}-park_s(\tilde{w})} t^{d-\dim(\tilde{w})}}
\label{final}
\end{equation}
where:
\begin{itemize}
\item $\tilde{w}$ is given by equation \ref{cell}.
    \item $d$ is the dimension of Hitchin's moduli space defined with general formula. \ref{modulidim}.
    \item The dimension statistic $-\dim(\tilde{w})$ for $\tilde{w}$ is computed as in equation \ref{celldim} and \ref{celldim1}
    \item The $park_s$ is defined using the map to parking function: $park(\tilde{w})=[a_1,\ldots, a_n]$, and $park_s(\tilde{w})=\sum a_i$.
\end{itemize}

\subsection{An example}

\textbf{Example 1}: Consider the root system of type $A_2$ with $\nu=4$ and $f=[1,1,1]$.
The set $\tilde{\Phi}_4$ is given by $\{\alpha_1+\delta, \alpha_2+\delta, -(\alpha_1+\alpha_2)+2\delta\}$.

Consider an affine Weyl group element $\tilde{w}=st_{m_1\alpha_1+m_2 \alpha_2}$. Its action on $\tilde{\Phi}_4$ is as follows:
\begin{align*}
&st_{m_1\alpha_1+m_2 \alpha_2} (\alpha_1+\delta)=s(\alpha_1)+(1-(2m_1-m_2))\delta, \nonumber\\
&st_{m_1\alpha_1+m_2 \alpha_2} (\alpha_2+\delta)=s(\alpha_2)+(1-(2m_2-m_1))\delta, \nonumber\\
&st_{m_1\alpha_1+m_2 \alpha_2} (-(\alpha_1+\alpha_2)+2\delta)=-s(\alpha_1+\alpha_2)+(2+(m_1+m_2))\delta.
\end{align*}

The finite Weyl group $W$ is generated by $\{1, s_1, s_2, s_1s_2, s_2s_1, s_1s_2s_1\}$.
Imposing the condition $\tilde{w}(\tilde{\Phi}_\nu)\subset \tilde{\Phi}_{+}$ yields the following constraints and solutions for $(m_1, m_2)$:
\begin{align*}
&s=1, & &2m_1-m_2\leq1,~\&~2m_2-m_1\leq 1~\&~2+m_1+m_2>0, & &(0,0),~(-1,0),~(0,-1),~(1,1), \nonumber\\
& s=s_1, & &2m_1-m_2<1,~\&~2m_2-m_1\leq 1~\&~2+m_1+m_2>0, & &(0,0),~(-1,0), \nonumber \\
& s=s_2, & &2m_1-m_2\leq1,~\&~2m_2-m_1< 1~\&~2+m_1+m_2>0, & &(0,0),~(0,-1), \nonumber\\
&s=s_1s_2, & &2m_1-m_2\leq1,~\&~2m_2-m_1< 1~\&~2+m_1+m_2\geq 0, & &(0,0),~(0,-1),~(-1,-1),\nonumber\\
&s=s_2s_1, & &2m_1-m_2<1,~\&~2m_2-m_1 \leq 1~\&~2+m_1+m_2\geq 0, & &(0,0),~(-1,0),~(-1,-1),\nonumber\\
&s=s_1s_2s_1, & &2m_1-m_2<1,~\&~2m_2-m_1< 1~\&~2+m_1+m_2\geq 0, & &(0,0),~(-1,-1).\nonumber
\end{align*}
There are a total of 16 solutions, corresponding to the following affine Weyl group elements:
\begin{align*}
&1,\ t_{-\alpha_1},\ t_{-\alpha_2},\ t_{\alpha_1+\alpha_2},\ s_1,\ s_1t_{-\alpha_1},\ s_2,\ s_2t_{-\alpha_2},\ s_1s_2,\ s_1s_2t_{-\alpha_2},\ s_1s_2t_{-\alpha_1-\alpha_2}, \nonumber\\
& s_2s_1,\ s_2s_1t_{-\alpha_1},\ s_2s_1t_{-\alpha_1-\alpha_2},\ s_1s_2s_1,\ s_1s_2s_1t_{-\alpha_1-\alpha_2}.
\end{align*}
Among these, the dominant elements are $1$, $t_{\alpha_1+\alpha_2}$, $s_1s_2t_{-\alpha_2}$, $s_2s_1t_{-\alpha_1}$, and $s_1s_2s_1t_{-\alpha_1-\alpha_2}$. The elements are partitioned into the following Weyl group orbits:
\begin{align}
&\text{orb}_1:\ (1,s_1,s_2,s_1s_2,s_2s_1,s_1s_2s_1), \nonumber\\
&\text{orb}_2:\ (t_{-\alpha_1},s_1t_{-\alpha_1}, s_2s_1t_{-\alpha_1}), \nonumber\\
&\text{orb}_3:\ (t_{-\alpha_2}, s_2t_{-\alpha_2},s_1s_2t_{-\alpha_2}), \nonumber\\
&\text{orb}_4:\ (s_1s_2t_{-\alpha_1-\alpha_2},s_2s_1t_{-\alpha_1-\alpha_2},s_1s_2s_1t_{-\alpha_1-\alpha_2}), \nonumber\\
&\text{orb}_5:\ (t_{\alpha_1+\alpha_2}).
\label{orbit}
\end{align}

\begin{table}[htp]
\caption{The coroots from the Anderson map and the corresponding parking function for $A_2$ case with $\nu=4$. The orbits of Weyl group elements are listed in \ref{orbit}.}
\begin{center}
\begin{tabular}{|c|c|c|c|} \hline
Orbits&Coroots & Coordinates & Parking \\ \hline
$orb_1$&$(\alpha_1, \alpha_2, \alpha_1+\alpha_2, -\alpha_1-\alpha_2, -\alpha_1, -\alpha_2)$ & $[1,-1,0]$ &$[2,0,1]$ \\ \hline
$orb_2$&$2\alpha_2+\alpha_1,~~-2\alpha_1-\alpha_2,~~\alpha_1-\alpha_2$ & $[1,1,-2]$ &$[0,0,1]$ \\ \hline
$orb_3$&$2\alpha_1+\alpha_2,~~-2\alpha_2-\alpha_1,~~\alpha_2-\alpha_1$ & $[2,-1,-1]$ & $[0,1,1]$\\ \hline
$orb_4$&$2\alpha_1+2\alpha_2,~~2\alpha_2, 2\alpha_1$ & $[2,0,-2]$ &$[0,2,0]$ \\ \hline
$orb_5$&$(0)$ & $[0,0,0]$ & $[0,0,0]$ \\ \hline
\end{tabular}
\end{center}
\label{a2park}
\end{table}%

The dimension $\dim(\tilde{w})$ is computed using Formula~\ref{celldim} \footnote{The set in \ref{celldim1} is empty for this case as $\Delta_P$ is empty.}. The following data is required for the computation of the cell dimension: the list of roots of height less than $4$ consists of $(\alpha_1, 0)$, $(\alpha_2, 0)$, $(\alpha_1+\alpha_2, 0)$, $(-\alpha_1, 1)$, $(-\alpha_2, 1)$, and $(-\alpha_1-\alpha_2, 1)$. The corresponding parking function is computed via the Anderson map; we list only one representative, with the others being permutations of it. The element $\tilde{w}_\nu = t_{-\alpha_1-\alpha_2}$ is used in Formula~\ref{anderson}.

Using Table~\ref{nu4data}, we obtain the bigraded polynomial from Formula~\ref{final} (with $n=3$, $\nu=4$, $\dim \mathcal{O}_f = 0$, and $d=3$):
\begin{align*}
&q^0(t^0 + 2t^2 + 2t + t^3) + q^2(2t^0 + t) + q(2t^0 + t) + q(2t + t^2) + q^3t^0.
\end{align*}
The $q,t$-polynomial in Schur expansion form, as shown in Table~\ref{examples}, is reproduced below:
\begin{align}
&[1]_{q,t} s_{[3]} + \left([2]_{q,t} + [3]_{1,t}\right) s_{[2,1]} + (qt + [4]_{q,t}) s_{[1,1,1]} \\
&= q^0 t^0 s_{[3]} + (q t^0 + q^0 t) s_{[2,1]} + (q^2 t^0 + q t + q^0 t^2) s_{[2,1]} + (q t + q^3 + q^2 t + q t^2 + t^3) s_{[1,1,1]}.
\label{nu4graded}
\end{align}
To match the above result, one substitutes the Schur polynomials with the dimensions of the corresponding irreducible representations of the symmetric group: $s_{[2,1]} \to 2$, $s_{[3]} \to 1$, and $s_{[1,1,1]} \to 1$.

\begin{table}[htp]
\centering
\begin{tabular}{|c|c|c|c|c|c|} \hline
Affine Weyl & $park_{s}$ & dim & Affine Weyl & $park_s$ & dim \\ \hline 
1 & 3 & 0 & $s_1$ & 3 & 1 \\ \hline 
$s_2$ & 3 & 1 & $s_1 s_2$ & 3 & 2 \\ \hline
$s_2 s_1$ & 3 & 2 & $s_1 s_2 s_1$ & 3 & 3 \\ \hline
& & & & & \\ \hline
$t_{-\alpha_1}$ & 1 & 3 & $s_1 t_{-\alpha_1}$ & 1 & 3 \\ \hline
$s_2 s_1 t_{-\alpha_1}$ & 1 & 2 & & & \\ \hline
& & & & & \\ \hline
$t_{-\alpha_2}$ & 2 & 3 & $s_2 t_{-\alpha_2}$ & 2 & 3 \\ \hline
$s_1 s_2 t_{-\alpha_2}$ & 2 & 2 & & & \\ \hline
& & & & & \\ \hline
$s_2 s_1 t_{-\alpha_1-\alpha_2}$ & 2 & 2 & $s_1 s_2 s_1 t_{-\alpha_1-\alpha_2}$ & 2 & 1 \\ \hline
$s_1 s_2 t_{-\alpha_1-\alpha_2}$ & 2 & 2 & & & \\ \hline
& & & & & \\ \hline
$t_{\alpha_1+\alpha_2}$ & 0 & 3 & & & \\ \hline
\end{tabular}
\caption{Affine Weyl group elements satisfying $\tilde{w}(\Phi_\nu) \subset \Phi^+$, with $\nu=4$ and $\mathfrak{g} = A_2$. This set is $W^\nu$. The column $park_s$ gives the sum of the entries of the corresponding parking function. The column $dim$ gives the dimension of the corresponding cell in the affine Springer fiber.}
\label{nu4data}
\end{table}%

To compare with the results from the previous section, we need to define a map from the affine Weyl group elements to the parking functions (equivalently, labeled Dyck paths). This is accomplished via the Anderson map, as shown in Table~\ref{a2park}.

\newpage
\textbf{Example 2}: Let $f = [2,1]$ with $\Delta_P = \{\alpha_1\}$, so $\tilde{n}_P$ is generated by
\begin{equation*}
\{\alpha + k\delta, k > 0, \alpha \in \Delta\} \cup \{\alpha_2, \alpha_1 + \alpha_2\}.
\end{equation*}
Following the same computation as in the previous example, we obtain the solutions:
\begin{align*}
&s = 1,\quad 2m_1 - m_2 < 1,\ \&\ 2m_2 - m_1 \leq 1,\ \&\ 2 + m_1 + m_2 > 0,\quad (0,0),\ (-1,0) \nonumber\\
&s = s_1,\quad 2m_1 - m_2 < 1,\ \&\ 2m_2 - m_1 \leq 1,\ \&\ 2 + m_1 + m_2 > 0,\quad (0,0),\ (-1,0) \nonumber\\
&s = s_2,\quad 2m_1 - m_2 \leq 1,\ \&\ 2m_2 - m_1 < 1,\ \&\ 2 + m_1 + m_2 > 0,\quad (0,0),\ (0,-1) \nonumber\\
&s = s_1 s_2,\quad 2m_1 - m_2 \leq 1,\ \&\ 2m_2 - m_1 < 1,\ \&\ 2 + m_1 + m_2 > 0,\quad (0,0),\ (0,-1)\nonumber\\
&s = s_2 s_1,\quad 2m_1 - m_2 < 1,\ \&\ 2m_2 - m_1 \leq 1,\ \&\ 2 + m_1 + m_2 \geq 0,\quad (0,0),\ (-1,-1)\nonumber\\
&s = s_1 s_2 s_1,\quad 2m_1 - m_2 < 1,\ \&\ 2m_2 - m_1 < 1,\ \&\ 2 + m_1 + m_2 \geq 0,\quad (0,0),\ (-1,-1).\nonumber
\end{align*}

We organize these into $\tilde{W}/s_1$ cosets:
\begin{align*}
&(1, s_1),\quad (s_1 t_{-\alpha_1}, t_{-\alpha_1}),\quad (s_2, s_1 s_2),\quad (s_2 t_{-\alpha_2}, s_1 s_2 t_{-\alpha_2}), \nonumber\\
&(s_2 s_1, s_1 s_2 s_1),\quad (s_1 s_2 s_1 t_{-\alpha_1-\alpha_2}, s_2 s_1 t_{-\alpha_1-\alpha_2}).
\end{align*}
There are a total of $6$ $s_1$-orbits, which matches the number given in Formula~\ref{tab:dimensions}. The dimension of the Hitchin moduli space is $d = 2$.

To compute the cell dimension, we require the following data. The list of roots of height less than $4$ is:
\begin{equation*}
(\alpha_1, 0),\ (\alpha_2, 0),\ (\alpha_1+\alpha_2, 0),\ (-\alpha_1, 1),\ (-\alpha_2, 1),\ (-\alpha_1-\alpha_2, 1).
\end{equation*}
For an affine Weyl group element $s t_q$, the conditions in Equation~\ref{celldim} are:
\begin{equation}
\begin{cases}
a):\ s(\alpha) \in \{\alpha_1, -\alpha_1, \alpha_2, \alpha_1+\alpha_2\}, & (k + (-q, \alpha) < 0, \\
b):\ s(\alpha) \in \{-\alpha_2, -\alpha_1-\alpha_2\}, & (k + (-q, \alpha) \leq 0.
\end{cases}
\end{equation}
The equation for the affine roots $\tilde{\alpha} = (\alpha, k)$ in Equation~\ref{celldim1} is:
\begin{equation*}
s(\alpha) \in \{\alpha_1, -\alpha_1\} \ \&\ k + (-q, \alpha) = 0 \ \&\ \text{height}(\tilde{\alpha}) \geq \nu.
\end{equation*}

The parking function is found using the same Anderson map as in the previous example; see Table~\ref{a2park}.

\begin{table}[htp]
\centering
\caption{Affine Weyl group elements satisfying $\tilde{w}(\Phi_\nu) \subset \tilde{n}_P$, where $P$ is generated by the simple root $\alpha_1$ of the $A_2$ Lie algebra.}
\begin{tabular}{|c|c|c|c|c|c|} \hline
Affine Weyl & $park_s$ & dim & Affine Weyl & $park_s$ & dim \\ \hline 
1 & 3 & 0 & $t_{-\alpha_1}$ & 1 & 2 \\ \hline 
$s_2$ & 3 & 1 & $s_2 t_{-\alpha_2}$ & 2 & 2 \\ \hline
$s_2 s_1$ & 3 & 2 & $s_2 s_1 t_{-\alpha_1-\alpha_2}$ & 2 & 1 \\ \hline
\end{tabular}
\label{nu4mini}
\end{table}%

Using the data in Table~\ref{nu4mini} and Formula~\ref{final} (with $d = 2$), the bigraded polynomial is:
\begin{equation*}
q^0(t^2 + t + t^0) + q t + q^2 t^0 + q t^0.
\end{equation*}
This agrees with the general Formula~\ref{eq:general_case} using the result from the previous example (with trivial regular singularity) and the Kostka numbers. The nonzero Kostka numbers are $K_{[[2,1],[2,1]]} = 1$ and $K_{[[3],[2,1]]} = 1$, so:
\begin{align*}
&C_{4,3}^{[2,1]} = q^0 t^0 K_{[[3],[2,1]]} + (q t^0 + q^0 t) K_{[[3],[2,1]]} + (q^2 t^0 + q t + q^0 t^2) K_{[[3],[2,1]]} \nonumber\\
&\quad = q^0 t^0 + (q t^0 + q^0 t) + (q^2 t^0 + q t + q^0 t^2).
\end{align*}

\newpage
\section{Verification with known results}

Although our proposed formula applies to the Hitchin system with fractional order poles, some cases are isomorphic to the Hitchin system with integral order poles.  

Consider the $sl_n$ wild Hitchin system with irregular singularities of integral order:
\begin{equation}
\Phi = \frac{T}{z^{r}} + \cdots
\label{pole}
\end{equation}
where $r\geq 2$ is a positive integer for irregular singularity, 
the mixed Hodge polynomial is computed in \cite{hausel2019arithmetic}. When $n=2$, the polynomial is given by :
\begin{align}
H(p,r,g) &= \frac{(q+1)^p t^{-2r} (q t+1)^{2g} (q^2 t+1)^{2g} (q t^2)^{2g+p+2r-2}}{(q^2-1)(q^2 t^2-1)} \\
&\quad + \frac{(q t+1)^{2g} (q^2 t^3+1)^{2g} (q t^2+1)^p}{(q^2 t^2-1)(q^2 t^4-1)} \\
&\quad - \frac{2^{p-1} (q t+1)^{4g} (q t^2)^{2g+p+r-2}}{(q-1)(q t^2-1)}
\label{sl2}
\end{align}
where $g$ is the genus, $p$ the total number of punctures (regular plus irregular), and $r = r_1 + \cdots + r_m$ is the sum of irregular pole orders \footnote{$r_i$ is defined as the total  irregular pole order minus one.}.

\subsection{$sl_2$ cases}
We consider two important special cases for $sl_2$ Hitchin system with integral pole order. We take genus $g=0$, and the formula becomes

\subsubsection{Case 1: One irregular puncture (order $r+2$)}
For a single irregular puncture with total pole order $r+2$ (see \ref{pole}), the mixed Hodge polynomial from \ref{sl2} ($p=1$, and $g=0$, order $r+2$) is
\begin{equation*}
\boxed{(A_1, A_{2r-1}):~ H(q,t) = \sum_{i=0}^{r-1}(q^i + q^{i+1} + \cdots + q^{2i}) t^{2i}}
\end{equation*}
The hyperkähler dimension of Hitchin's moduli space is $d = r-1$, yielding the combinatorial polynomial:
\begin{align}
C(q,t) &= H(q,(qt)^{-1/2}) t^d = H(q,(qt)^{-1/2}) t^{r-1} \nonumber \\
&= \sum_{i=0}^{r-1}(q^i + q^{i+1} + \cdots + q^{2i}) q^{-i} t^{r-i-1} \nonumber \\
&= \sum_{i=0}^{r-1}(1 + q + \cdots + q^{i-1} + q^i) t^{r-i-1}
\label{a1an}
\end{align}
This polynomial has $\frac{r(r+1)}{2}$ terms.

\subsubsection{Case 2: One irregular and one regular puncture}
For one irregular puncture of order $r-1$ and one regular puncture, the polynomial from \ref{sl2} (($p=2,~g=0,~order~r-1$)) is
\begin{equation*}
\boxed{(A_1, D_{2r-2}):~ H(q,t) = \sum_{i=0}^{r-2}(2q^i + 2q^{i+1} + \cdots + 2q^{2i-1} + q^{2i}) t^{2i}}
\end{equation*}
The corresponding combinatorial polynomial  is:
\begin{align}
C(q,t) &= H(q,(qt)^{-1/2}) t^d = H(q,(qt)^{-1/2}) t^{r-2} \nonumber \\
&= \sum_{i=0}^{r-2}(2q^i + 2q^{i+1} + \cdots + 2q^{2i-1} + q^{2i}) q^{-i} t^{r-i-2} \nonumber \\
&= \sum_{i=0}^{r-2}(2 + 2q + \cdots + 2q^{i-1} + q^i) t^{r-i-2}
\label{a1dn}
\end{align}
with hyperkähler dimension $d = r-2$. This polynomial contains $(r-1)^2$ terms.

\subsection{Verification using our formula}
These theories can be engineered using specific regular singularities \cite{Song:2017oew} of higher rank Hitchin system fitting in \ref{fig:singularities}. We need to do computation of rational parking functions for $m = N-1, N+1$ and $n = N$, see table. \ref{tab:correspondence}.

\begin{table}[htp]
\centering
\caption{The ramified Hitchin system corresponding to unramified $sl_2$ Hitchin system.}
\begin{tabular}{|c|c|}
\hline
$(A_1, A_{2N-1})$ & $n=N$, $m=N+1$, $f=[N-1,1]$ \\ \hline
$(A_1, D_{2N-2})$ & $n=N$, $m=N-1$, $f=[N-2,1^2]$ \\ \hline
\end{tabular}
\label{tab:correspondence}
\end{table}

The combinatorial polynomial for the first case (with regular singularities $f=[1,\ldots, 1]$) is:
\begin{equation*}
C_{N+1,N}(q,t) = \cdots + 
\begin{bmatrix}
0 & 1 & 1 & 1 & 1 \\
1 & 1 & 1 & 1 & 0 \\
1 & 1 & 1 & 0 & 0 \\
1 & 1 & 0 & 0 & 0 \\
1 & 0 & 0 & 0 & 0
\end{bmatrix}_{N \times N} s_{[N-1,1]} + [1]s_{[N]}
\end{equation*}
Only one Dyck path with pattern $NN\cdots NN$ contributes to the term $s_{[N]}$. The coefficient of $s_{[N-1,1]}$ receives contribution from:
\begin{itemize}
\item The Dyck path $N^2N\cdots NN$
\item The Kostka number $K_{[2,\ldots,1],[1,\ldots,1]} = N-1$ from $NN\cdots NN$
\end{itemize}
The notation $\ldots N^i \ldots$ implies that there is a column with $i$ number of vertical steps in the Dyck path.

Using our general formula and the above rational parking function, we find
\begin{equation*}
C^{[N-1,1]}_{N+1,N}(q,t) = 
\begin{bmatrix}
0 & 1 & 1 & 1 & 1 \\
1 & 1 & 1 & 1 & 0 \\
1 & 1 & 1 & 0 & 0 \\
1 & 1 & 0 & 0 & 0 \\
1 & 0 & 0 & 0 & 0
\end{bmatrix}
+ [1] = 
\begin{bmatrix}
1 & 1 & 1 & 1 & 1 \\
1 & 1 & 1 & 1 & 0 \\
1 & 1 & 1 & 0 & 0 \\
1 & 1 & 0 & 0 & 0 \\
1 & 0 & 0 & 0 & 0
\end{bmatrix}_{N \times N}
\end{equation*}
This matches equation \eqref{a1an}.

For the second case, the polynomial from the rational parking function is:
\begin{equation*}
C_{N-1,N}(q,t) = \cdots + 
\begin{bmatrix}
0 & 1 & 1 & 1 & 1 \\
1 & 1 & 1 & 1 & 0 \\
1 & 1 & 1 & 0 & 0 \\
1 & 1 & 0 & 0 & 0 \\
1 & 0 & 0 & 0 & 0
\end{bmatrix}_{N-1 \times N-1} s_{[N-2,1,1]} + 
\begin{bmatrix}
0 & 1 & 1 & 1 \\
1 & 1 & 1 & 0 \\
1 & 1 & 0 & 0 \\
1 & 0 & 0 & 0
\end{bmatrix}_{N-2 \times N-2} s_{[N-2,2]} + [1]s_{[N-1,1]}
\end{equation*}
The term $s_{[N-1,1]}$ receives contributions from: Exactly one Dyck path with the pattern $N^2N\ldots NN$.
The coefficient of $s_{[N-2,2]}$ receives following contributions: a) The Dyck path $N^2N^2\ldots NN$; 
b): An additional $N-3$ contributions from the path $N^2N\ldots NN$.
The coefficient of $s_{[N-2,1,1]}$ receives following contributions: a) The path $N^3N\ldots NN$; b): $N-2$ terms from $N^2N\ldots NN$ (counted by the Kostka number $K_{[3,\ldots,1],[2,\ldots,1]} = N-2$); c) One additional term from $N^2N^2\ldots NN$.

The relevant Kostka numbers are:
\begin{align*}
K_{[N-1,1],[N-2,1,1]} &= 2 \\
K_{[N-2,2],[N-2,1,1]} &= 1 \\
K_{[N-2,1,1],[N-2,1,1]} &= 1
\end{align*}
This gives:
\begin{align*}
C^{[N-2,1,1]}_{N-1,N}(q,t) &= 
\begin{bmatrix}
0 & 1 & 1 & 1 & 1 \\
1 & 1 & 1 & 1 & 0 \\
1 & 1 & 1 & 0 & 0 \\
1 & 1 & 0 & 0 & 0 \\
1 & 0 & 0 & 0 & 0
\end{bmatrix}_{N-1 \times N-1} + 
\begin{bmatrix}
0 & 1 & 1 & 1 \\
1 & 1 & 1 & 0 \\
1 & 1 & 0 & 0 \\
1 & 0 & 0 & 0
\end{bmatrix}_{N-2 \times N-2} + 2[1] \\
&= 
\begin{bmatrix}
2 & 2 & 2 & 2 & 1 \\
2 & 2 & 2 & 1 & 0 \\
2 & 2 & 1 & 0 & 0 \\
2 & 1 & 0 & 0 & 0 \\
1 & 0 & 0 & 0 & 0
\end{bmatrix}_{N-1 \times N-1}
\end{align*}
which agrees with equation \eqref{a1dn}.

\section{Conclusion}
We propose a bigraded polynomial for ramified wild Hitchin systems of type $A$, conjecturing that it coincides with both the polynomial arising from the perverse filtration of the Hitchin fibration and the weight filtration from the mixed Hodge structure. This bigraded polynomial can be computed either from rational parking functions or using results from affine Springer fibers. It would be interesting to verify our conjectures by studying the perverse filtration of the Hitchin fibration or by examining the geometry of the corresponding character variety. Some cases have already been computed in \cite{hausel2019arithmetic,szabo2021perversity}, and we hope that the general structure proposed in this paper will facilitate further geometric computations, thereby providing insights for the $P=W$ conjecture in this more general context. The bigraded polynomial is also related to the double affine Hecke algebra (DAHA) \cite{berest2003finite}, and it would be interesting to verify our results by investigating the structure of DAHA.

Generalizing our results to Hitchin systems defined using other types of Lie algebras presents an interesting direction for future work. The cohomological grading can be readily computed using Formulas~\ref{celldim} and~\ref{celldim1}, which are valid in all cases. However, determining the other grading is considerably more challenging. The type $A$ case appears to be special, as it depends only on the root vector of the affine Weyl group element parameterizing the cell of the affine Springer fiber.

The bigraded rational parking function is a central object in algebraic combinatorics \cite{haglund2008q}. We hope that the geometry of the Hitchin moduli space will contribute to a deeper understanding of these objects, particularly by revealing geometric origins for the gradings used to define graded rational parking functions. For example, the remarkable $q,t$ symmetric properties of rational parking function \cite{armstrong2016rational} might be explained geometrically.

Our work is motivated by the interpretation of Hitchin moduli spaces as Coulomb branch solutions of 4D $\mathcal{N}=2$ theories. The bigraded polynomials arising from the Coulomb branch have interesting physical implications, which will be explored in future work. These polynomials actually apply to Coulomb branches of arbitrary 4D $\mathcal{N}=2$ theories. Through circle compactification, they can be computed either via the fibration structure of the 4D theory or through the geometry of the corresponding 3D $\mathcal{N}=4$ theories, which are related to character varieties. This suggests that the $P=W$ phenomenon may hold generally for geometries arising from \textbf{any} 4D $\mathcal{N}=2$ theory, with Hitchin moduli spaces representing a special case.

\section*{Acknowledgements}
This work is supported by the National Key Research and Development Program of China (NO.
2020YFA0713000).

\newpage
\appendix

\section{Symmetric polynomials}
\label{symmetric}

The \textbf{Schur polynomial} for a partition $\lambda \in P_n$ and $k$ variables  is defined as 
\begin{equation*}
S_\lambda={|x_j^{\lambda_i+k-i}| \over \Delta} 
\end{equation*}
Here $\Delta=\prod_{i<j}(x_i-x_j)$ is the discriminant, and $|a_{i,j}|$ denotes the discriminant of the $k\times k$ matrix with entry $a_{i,j}$.  

By taking $n=2, k=2$, the Schur polynomial is then
\begin{equation*}
s_{(2)}(x_1, x_2)=x_1^2+x_1 x_2+x_2^2,~~~s_{(1,1)}(x_1, x_2)=x_1 x_2
\end{equation*}
By taking $n=3, k=3$, the Schur polynomial is then
\begin{align*}
& s_{(3)}(x_1, x_2, x_3) = x_1^3 + x_1^2x_2 + x_1^2x_3 + x_1x_2^2 + x_1x_2x_3 + x_1x_3^2 + x_2^3 + x_2^2x_3 + x_2x_3^2 + x_3^3 \nonumber\\
&s_{(2,1)}(x_1, x_2, x_3) = x_1^2x_2 + x_1^2x_3 + x_1x_2^2 + 2x_1x_2x_3 + x_1x_3^2 + x_2^2x_3 + x_2x_3^2, \nonumber\\
&s_{(1,1,1)}(x_1, x_2, x_3) = x_1x_2x_3 
\end{align*}

A polynomial \( Q \in \mathbb{Q}[x_1, \dots, x_n] \) is \textbf{quasi-symmetric} if for any two sequences of exponents \( \alpha = (\alpha_1, \dots, \alpha_k) \) and \( \beta = (\beta_1, \dots, \beta_k) \), the coefficients of the monomials \( x_{i_1}^{\alpha_1} \cdots x_{i_k}^{\alpha_k} \) and \( x_{j_1}^{\beta_1} \cdots x_{j_k}^{\beta_k} \) are equal whenever \( i_1 < \cdots < i_k \) and \( j_1 < \cdots < j_k \).

For a composition \( \alpha = (\alpha_1, \dots, \alpha_k) \), the \textbf{fundamental quasi-symmetric polynomial} \( F_\alpha \) is defined as:
\[
F_\alpha(x_1, \dots, x_n) = \sum_{\substack{1 \leq i_1 \leq \cdots \leq i_n \\ i_j < i_{j+1} \text{ if } j \in D(\alpha)}} x_{i_1} \cdots x_{i_n},
\]
where \( D(\alpha) \) is the descent set of \( \alpha \).

A Schur polynomial \( s_\lambda \) decomposes into fundamental quasi-symmetric polynomials as:
\[
s_\lambda(x_1, \dots, x_n) = \sum_{T \in \text{SYT}(\lambda)} F_{\text{des}(T)}(x_1, \dots, x_n),
\]
where:
\begin{itemize}
\item \( \text{SYT}(\lambda) \) is the set of standard Young tableaux of shape \( \lambda \) \footnote{The standard Young tableaux is defined by adding labels to Young Tableaux so that the labels on the rows 
increase strictly from left to right, and the labels on columns increase strictly from top to bottom.}.
\item \( \text{des}(T) \) is the descent set from \( T \).
\end{itemize}

\textbf{Example}:~[Partition \( \lambda = (2,1) \)]
The standard Young tableaux (SYT) of shape \( (2,1) \) are:
\[
T_1 = young(12,3) \quad \text{(descent set } \{2\})
\]
\[
T_2 = young(13,2) \quad \text{(descent set } \{1\})
\]
Thus, the Schur polynomial decomposes as:
\[
s_{(2,1)}(x_1, x_2, x_3) = F_{2} + F_{1},
\]

The list  of fundamental quasi-symmetric functions of $S_3$ group are
\begin{align*}
& F_{\emptyset}=s_{(3)}\nonumber\\
&F_{1}= x_1x_2^2 + x_1x_2x_3 + x_1x_3^2+x_2x_3^2, \nonumber\\
&F_{2} = x_1^2x_2 + x_1^2x_3 +x_1x_2x_3+ x_2^2x_3 \nonumber\\
&F_{12}=s_{(1,1,1)} 
\end{align*}
and so we have $s_{(2,1)}=F_{1}+F_{2}$.  Therefore, for the Dyck path in figure. \ref{fig:dyck_path}, the Schur expansion is $s_{(2,1)}+s_{(1,1,1)}$ as it consists of three fundamental quasi-symmetric polynomials.

\section{Rational parking function with $m=5, n=3$.}
We give the combinatorial data for the rational parking function with $m=5, n=3$. The rational parking function is 
\begin{equation*}
C_{5,3}(q,t)=\left[\begin{array}{ccccc} 
0&0&0&0&1\\
0&0&1&1&0\\
0&1&1&0&0\\
0&1&0&0&0\\
1&0&0&0&0 \end{array} \right]s_{[1,1,1]}+\left[\begin{array}{cccc} 
0&0&1&1\\
0&2&1&0\\
1&1&0&0\\
1&0&0&0\\ \end{array} \right]s_{[2,1]}+\left[\begin{array}{cc} 
0&1\\
1&0\\
 \end{array} \right]s_{[3]}
\end{equation*}
We used the decomposition of the Schur expansion: 
\begin{align*}
& NNN:~~s_{[1,1,1]}\oplus 2s_{[2,1]}  \oplus s_{[3]} \\
& N^2 N:~~s_{[1,1,1]} \oplus s_{[2,1]} \\
&N^3:~~s_{[1,1,1]}
\end{align*}
where $N^i$ indicates the vertical path has a column with $i$ step.

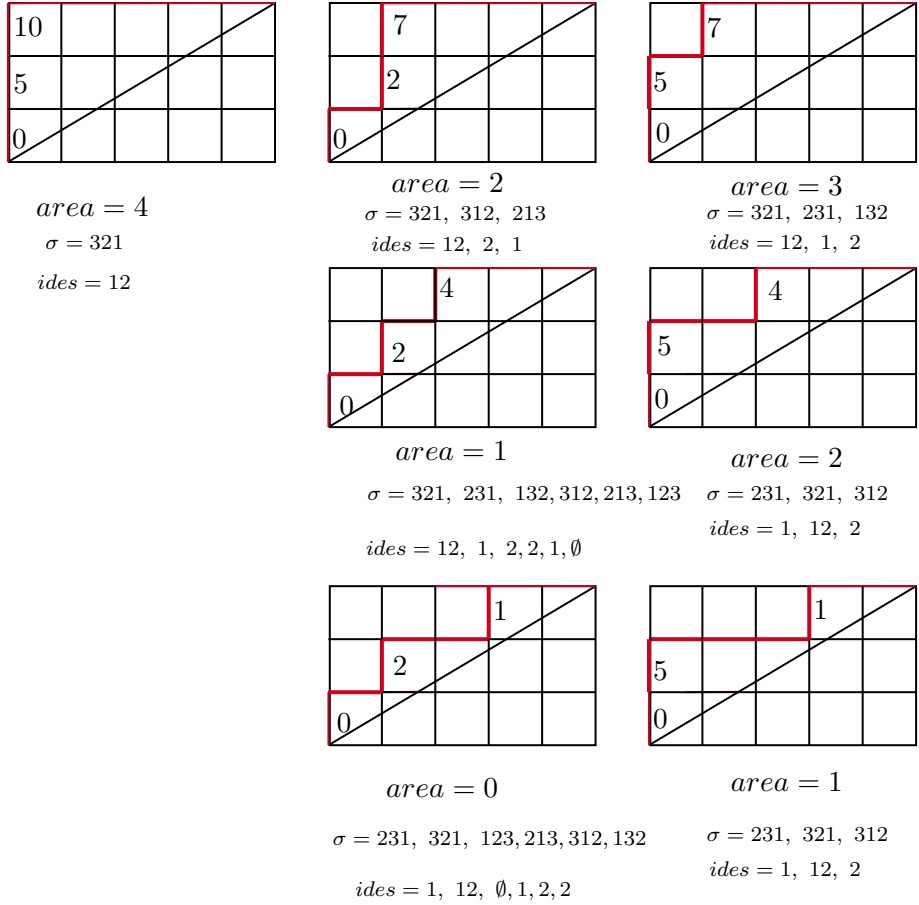
\begin{figure}
\begin{center}

\tikzset{every picture/.style={line width=0.75pt}} 

\begin{tikzpicture}[x=0.50pt,y=0.50pt,yscale=-1,xscale=1]

\draw   (121,80) -- (320,80) -- (320,200) -- (121,200) -- cycle ;
\draw    (120,200) -- (320,80) ;
\draw [color={rgb, 255:red, 208; green, 2; blue, 27 }  ,draw opacity=1 ]   (120,80.28) -- (160,80) ;
\draw [color={rgb, 255:red, 208; green, 2; blue, 27 }  ,draw opacity=1 ]   (120,160) -- (120,200) ;
\draw [color={rgb, 255:red, 208; green, 2; blue, 27 }  ,draw opacity=1 ]   (120,120) -- (120,160) ;
\draw [color={rgb, 255:red, 208; green, 2; blue, 27 }  ,draw opacity=1 ]   (160,80.28) -- (320,80) ;
\draw    (120,120) -- (320,120) ;
\draw    (280,200) -- (280,80) ;
\draw    (240,200) -- (240,80) ;
\draw    (200,200) -- (200,80) ;
\draw    (160,200) -- (160,120) ;
\draw    (160,120) -- (160,80) ;
\draw    (120,160) -- (320,160) ;
\draw [color={rgb, 255:red, 208; green, 2; blue, 27 }  ,draw opacity=1 ]   (120,80) -- (120,120) ;
\draw   (361,80) -- (560,80) -- (560,200) -- (361,200) -- cycle ;
\draw    (360,200) -- (560,80) ;
\draw [color={rgb, 255:red, 208; green, 2; blue, 27 }  ,draw opacity=1 ]   (387,160) -- (427,159.72) ;
\draw [color={rgb, 255:red, 208; green, 2; blue, 27 }  ,draw opacity=1 ]   (360,160) -- (360,200) ;
\draw [color={rgb, 255:red, 208; green, 2; blue, 27 }  ,draw opacity=1 ][line width=1.5]    (400,120) -- (400,160) ;
\draw [color={rgb, 255:red, 208; green, 2; blue, 27 }  ,draw opacity=1 ]   (400,80.28) -- (560,80) ;
\draw    (360,120) -- (560,120) ;
\draw    (520,200) -- (520,80) ;
\draw    (480,200) -- (480,80) ;
\draw    (440,200) -- (440,80) ;
\draw    (400,200) -- (400,160) ;
\draw    (640,160) -- (640,120) ;
\draw    (360,160) -- (560,160) ;
\draw [color={rgb, 255:red, 208; green, 2; blue, 27 }  ,draw opacity=1 ][line width=1.5]    (400,320) -- (440,320) ;
\draw [color={rgb, 255:red, 208; green, 2; blue, 27 }  ,draw opacity=1 ][line width=1.5]    (400,80) -- (400,120) ;
\draw [color={rgb, 255:red, 208; green, 2; blue, 27 }  ,draw opacity=1 ][line width=1.5]    (360,160) -- (400,160) ;
\draw   (601,80) -- (800,80) -- (800,200) -- (601,200) -- cycle ;
\draw    (600,200) -- (800,80) ;
\draw [color={rgb, 255:red, 208; green, 2; blue, 27 }  ,draw opacity=1 ]   (627,160) -- (667,159.72) ;
\draw [color={rgb, 255:red, 208; green, 2; blue, 27 }  ,draw opacity=1 ]   (600,160) -- (600,200) ;
\draw [color={rgb, 255:red, 208; green, 2; blue, 27 }  ,draw opacity=1 ][line width=1.5]    (440,280) -- (440,320) ;
\draw [color={rgb, 255:red, 208; green, 2; blue, 27 }  ,draw opacity=1 ]   (640,80.28) -- (800,80) ;
\draw    (640,120) -- (800,120) ;
\draw    (760,200) -- (760,80) ;
\draw    (720,200) -- (720,80) ;
\draw    (680,200) -- (680,80) ;
\draw    (640,200) -- (640,160) ;
\draw    (600,160) -- (800,160) ;
\draw [color={rgb, 255:red, 208; green, 2; blue, 27 }  ,draw opacity=1 ][line width=1.5]    (640,80) -- (640,120) ;
\draw [color={rgb, 255:red, 208; green, 2; blue, 27 }  ,draw opacity=1 ][line width=1.5]    (600,120) -- (600,160) ;
\draw   (361,280) -- (560,280) -- (560,400) -- (361,400) -- cycle ;
\draw    (360,400) -- (560,280) ;
\draw [color={rgb, 255:red, 208; green, 2; blue, 27 }  ,draw opacity=1 ]   (387,360) -- (427,359.72) ;
\draw [color={rgb, 255:red, 208; green, 2; blue, 27 }  ,draw opacity=1 ]   (360,360) -- (360,400) ;
\draw [color={rgb, 255:red, 208; green, 2; blue, 27 }  ,draw opacity=1 ][line width=1.5]    (400,320) -- (400,360) ;
\draw [color={rgb, 255:red, 208; green, 2; blue, 27 }  ,draw opacity=1 ]   (440,280) -- (560,280) ;
\draw    (360,320) -- (560,320) ;
\draw    (520,400) -- (520,280) ;
\draw    (480,400) -- (480,280) ;
\draw    (440,400) -- (440,280) ;
\draw    (400,400) -- (400,360) ;
\draw    (360,360) -- (560,360) ;
\draw [color={rgb, 255:red, 208; green, 2; blue, 27 }  ,draw opacity=1 ][line width=1.5]    (600,320) -- (640,320) ;
\draw [color={rgb, 255:red, 208; green, 2; blue, 27 }  ,draw opacity=1 ][line width=1.5]    (360,360) -- (400,360) ;
\draw    (400,320) -- (400,280) ;
\draw    (640,360) -- (640,320) ;
\draw   (601,280) -- (800,280) -- (800,400) -- (601,400) -- cycle ;
\draw    (600,400) -- (800,280) ;
\draw [color={rgb, 255:red, 208; green, 2; blue, 27 }  ,draw opacity=1 ]   (627,360) -- (667,359.72) ;
\draw [color={rgb, 255:red, 208; green, 2; blue, 27 }  ,draw opacity=1 ]   (600,360) -- (600,400) ;
\draw [color={rgb, 255:red, 208; green, 2; blue, 27 }  ,draw opacity=1 ]   (680,280) -- (800,280) ;
\draw    (680,320) -- (800,320) ;
\draw    (760,400) -- (760,280) ;
\draw    (720,400) -- (720,280) ;
\draw    (680,400) -- (680,280) ;
\draw    (640,400) -- (640,360) ;
\draw    (600,360) -- (800,360) ;
\draw [color={rgb, 255:red, 208; green, 2; blue, 27 }  ,draw opacity=1 ][line width=1.5]    (600,320) -- (600,360) ;
\draw [color={rgb, 255:red, 208; green, 2; blue, 27 }  ,draw opacity=1 ][line width=1.5]    (680,320) -- (640,320) ;
\draw [color={rgb, 255:red, 208; green, 2; blue, 27 }  ,draw opacity=1 ][line width=1.5]    (680,280) -- (680,320) ;
\draw [color={rgb, 255:red, 208; green, 2; blue, 27 }  ,draw opacity=1 ][line width=1.5]    (600,560) -- (640,560) ;
\draw    (640,600) -- (640,560) ;
\draw   (601,520) -- (800,520) -- (800,640) -- (601,640) -- cycle ;
\draw    (600,640) -- (800,520) ;
\draw [color={rgb, 255:red, 208; green, 2; blue, 27 }  ,draw opacity=1 ]   (627,600) -- (667,599.72) ;
\draw [color={rgb, 255:red, 208; green, 2; blue, 27 }  ,draw opacity=1 ]   (600,600) -- (600,640) ;
\draw [color={rgb, 255:red, 208; green, 2; blue, 27 }  ,draw opacity=1 ]   (720,520) -- (800,520) ;
\draw    (720,560) -- (800,560) ;
\draw    (760,640) -- (760,520) ;
\draw    (720,640) -- (720,520) ;
\draw    (680,640) -- (680,520) ;
\draw    (640,640) -- (640,600) ;
\draw    (600,600) -- (800,600) ;
\draw [color={rgb, 255:red, 208; green, 2; blue, 27 }  ,draw opacity=1 ][line width=1.5]    (600,560) -- (600,600) ;
\draw [color={rgb, 255:red, 208; green, 2; blue, 27 }  ,draw opacity=1 ][line width=1.5]    (680,560) -- (640,560) ;
\draw [color={rgb, 255:red, 208; green, 2; blue, 27 }  ,draw opacity=1 ][line width=1.5]    (720,560) -- (680,560) ;
\draw [color={rgb, 255:red, 208; green, 2; blue, 27 }  ,draw opacity=1 ][line width=1.5]    (720,520) -- (720,560) ;
\draw    (640,560) -- (640,520) ;
\draw [color={rgb, 255:red, 208; green, 2; blue, 27 }  ,draw opacity=1 ][line width=1.5]    (640,120) -- (600,120) ;
\draw    (640,320) -- (640,280) ;
\draw [color={rgb, 255:red, 208; green, 2; blue, 27 }  ,draw opacity=1 ][line width=1.5]    (400,560) -- (458,560) -- (480,560) ;
\draw [color={rgb, 255:red, 208; green, 2; blue, 27 }  ,draw opacity=1 ][line width=1.5]    (480,520) -- (480,560) ;
\draw   (361,520) -- (560,520) -- (560,640) -- (361,640) -- cycle ;
\draw    (360,640) -- (560,520) ;
\draw [color={rgb, 255:red, 208; green, 2; blue, 27 }  ,draw opacity=1 ]   (387,600) -- (427,599.72) ;
\draw [color={rgb, 255:red, 208; green, 2; blue, 27 }  ,draw opacity=1 ]   (360,600) -- (360,640) ;
\draw [color={rgb, 255:red, 208; green, 2; blue, 27 }  ,draw opacity=1 ][line width=1.5]    (400,560) -- (400,600) ;
\draw [color={rgb, 255:red, 208; green, 2; blue, 27 }  ,draw opacity=1 ]   (440,520) -- (560,520) ;
\draw    (480,560) -- (560,560) ;
\draw    (520,640) -- (520,520) ;
\draw    (480,640) -- (480,560) ;
\draw    (440,640) -- (440,560) ;
\draw    (400,640) -- (400,600) ;
\draw    (360,600) -- (560,600) ;
\draw [color={rgb, 255:red, 208; green, 2; blue, 27 }  ,draw opacity=1 ][line width=1.5]    (360,600) -- (400,600) ;
\draw    (400,560) -- (400,520) ;
\draw    (360,560) -- (400,560) ;
\draw    (440,560) -- (440,520) ;

\draw (121,173.4) node [anchor=north west][inner sep=0.75pt]    {$0$};
\draw (122,131.4) node [anchor=north west][inner sep=0.75pt]    {$5$};
\draw (122,88.4) node [anchor=north west][inner sep=0.75pt]    {$10$};
\draw (139,224.4) node [anchor=north west][inner sep=0.75pt]    {$area=4$};
\draw (405,206.4) node [anchor=north west][inner sep=0.75pt]    {$area=2$};
\draw (408,408.4) node [anchor=north west][inner sep=0.75pt]    {$area=1$};
\draw (401,662.4) node [anchor=north west][inner sep=0.75pt]    {$area=0$};
\draw (659,208.4) node [anchor=north west][inner sep=0.75pt]    {$area=3$};
\draw (659,413.4) node [anchor=north west][inner sep=0.75pt]    {$area=2$};
\draw (659,658.4) node [anchor=north west][inner sep=0.75pt]    {$area=1$};
\draw (361,173.4) node [anchor=north west][inner sep=0.75pt]    {$0$};
\draw (401,128.4) node [anchor=north west][inner sep=0.75pt]    {$2$};
\draw (406,87.4) node [anchor=north west][inner sep=0.75pt]    {$7$};
\draw (366,374.4) node [anchor=north west][inner sep=0.75pt]    {$0$};
\draw (405,334.4) node [anchor=north west][inner sep=0.75pt]    {$2$};
\draw (441,285.4) node [anchor=north west][inner sep=0.75pt]    {$4$};
\draw (364,613.4) node [anchor=north west][inner sep=0.75pt]    {$0$};
\draw (406,570.4) node [anchor=north west][inner sep=0.75pt]    {$2$};
\draw (481,529.4) node [anchor=north west][inner sep=0.75pt]    {$1$};
\draw (601,610.4) node [anchor=north west][inner sep=0.75pt]    {$0$};
\draw (601,574.4) node [anchor=north west][inner sep=0.75pt]    {$5$};
\draw (721,528.4) node [anchor=north west][inner sep=0.75pt]    {$1$};
\draw (602,369.4) node [anchor=north west][inner sep=0.75pt]    {$0$};
\draw (604,329.4) node [anchor=north west][inner sep=0.75pt]    {$5$};
\draw (687,287.4) node [anchor=north west][inner sep=0.75pt]    {$4$};
\draw (603,168.4) node [anchor=north west][inner sep=0.75pt]    {$0$};
\draw (601,130.4) node [anchor=north west][inner sep=0.75pt]    {$5$};
\draw (641,89.4) node [anchor=north west][inner sep=0.75pt]    {$7$};
\draw (385,231.4) node [anchor=north west][inner sep=0.75pt]  [font=\scriptsize]  {$\sigma =321,\ 312,\ 213$};
\draw (390,253.4) node [anchor=north west][inner sep=0.75pt]  [font=\scriptsize]  {$ides=12,\ 2,\ 1$};
\draw (643,252.4) node [anchor=north west][inner sep=0.75pt]  [font=\scriptsize]  {$ides=12,\ 1,\ 2$};
\draw (641,442.4) node [anchor=north west][inner sep=0.75pt]  [font=\scriptsize]  {$\sigma =231,\ 321,\ 312$};
\draw (643,468.4) node [anchor=north west][inner sep=0.75pt]  [font=\scriptsize]  {$ides=1,\ 12,\ 2$};
\draw (641,699.4) node [anchor=north west][inner sep=0.75pt]  [font=\scriptsize]  {$\sigma =231,\ 321,\ 312$};
\draw (643,725.4) node [anchor=north west][inner sep=0.75pt]  [font=\scriptsize]  {$ides=1,\ 12,\ 2$};
\draw (641,230.4) node [anchor=north west][inner sep=0.75pt]  [font=\scriptsize]  {$\sigma =321,\ 231,\ 132$};
\draw (387,442.4) node [anchor=north west][inner sep=0.75pt]  [font=\scriptsize]  {$\sigma =321,\ 231,\ 132,312,213,123$};
\draw (146,253.4) node [anchor=north west][inner sep=0.75pt]  [font=\scriptsize]  {$\sigma =321$};
\draw (140,282.4) node [anchor=north west][inner sep=0.75pt]  [font=\scriptsize]  {$ides=12$};
\draw (386,482.4) node [anchor=north west][inner sep=0.75pt]  [font=\scriptsize]  {$ides=12,\ 1,\ 2,2,1,\emptyset $};
\draw (361,702.4) node [anchor=north west][inner sep=0.75pt]  [font=\scriptsize]  {$\sigma =231,\ 321,\ 123,213,312,132$};
\draw (378,739.4) node [anchor=north west][inner sep=0.75pt]  [font=\scriptsize]  {$ides=1,\ 12,\ \emptyset ,1,2,2$};

\end{tikzpicture}

\end{center}
\caption{The combinatorial data for $5/3$ rational parking function. The numbers on the path is the rank function computed using formula. \ref{rank}.}
\label{append}
\end{figure}

\newpage

\bibliographystyle{JHEP}

\bibliography{ADhigher}

\providecommand{\href}[2]{#2}\begingroup\raggedright\begin{thebibliography}{10}

\bibitem{Xie:2012hs}
D.~Xie, {\it {General Argyres-Douglas Theory}},  {\em JHEP} {\bf 1301} (2013)
  100, [\href{http://xxx.lanl.gov/abs/1204.2270}{{\tt arXiv:1204.2270}}].

\bibitem{goresky2006purity}
M.~Goresky, R.~Kottwitz, and R.~MacPherson, {\it Purity of equivalued affine
  springer fibers},  {\em Representation Theory of the American Mathematical
  Society} {\bf 10} (2006), no.~6 130--146.

\bibitem{goresky2004homology}
M.~Goresky, R.~Kottwitz, and R.~MacPherson, {\it Homology of affine springer
  fibers in the unramified case},  {\em Duke Math. J.} {\bf 121} (2004), no.~1
  509--561.

\bibitem{varagnolo2009finite}
M.~Varagnolo and E.~Vasserot, {\it Finite-dimensional representations of daha
  and affine springer fibers: The spherical case},  {\em Duke Math. J.} {\bf
  146} (2009), no.~1 439--540.

\bibitem{oblomkov2016geometric}
A.~Oblomkov and Z.~Yun, {\it Geometric representations of graded and rational
  cherednik algebras},  {\em Advances in Mathematics} {\bf 292} (2016)
  601--706.

\bibitem{shan2023mirror}
P.~Shan, D.~Xie, and W.~Yan, {\it Mirror symmetry for circle compactified 4d
  $n= 2$ scfts},  {\em arXiv preprint arXiv:2306.15214} (2023).

\bibitem{shan2024modularity}
P.~Shan, D.~Xie, and W.~Yan, {\it Modularity for w-algebras and affine springer
  fibres},  {\em arXiv preprint arXiv:2404.00760} (2024).

\bibitem{de2010perverse}
M.~A.~A. de~Cataldo and L.~Migliorini, {\it The perverse filtration and the
  lefschetz hyperplane theorem},  {\em Annals of mathematics} (2010)
  2089--2113.

\bibitem{de2012topology}
M.~A.~A. de~Cataldo, T.~Hausel, and L.~Migliorini, {\it Topology of hitchin
  systems and hodge theory of character varieties: the case a 1},  {\em Annals
  of Mathematics} (2012) 1329--1407.

\bibitem{hausel2022p}
T.~Hausel, A.~Mellit, A.~Minets, and O.~Schiffmann, {\it $ p= w $ via $ h_2$},
  {\em arXiv e-prints} (2022) arXiv--2209.

\bibitem{maulik2024p}
D.~Maulik and J.~Shen, {\it The p=w conjecture for gl\_n},  {\em Annals of
  Mathematics} {\bf 200} (2024), no.~2 529--556.

\bibitem{garsia1996remarkable}
A.~M. Garsia and M.~Haiman, {\it A remarkable q, t-catalan sequence and
  q-lagrange inversion},  {\em Journal of Algebraic Combinatorics} {\bf 5}
  (1996), no.~3 191--244.

\bibitem{hikita2014affine}
T.~Hikita, {\it Affine springer fibers of type a and combinatorics of diagonal
  coinvariants},  {\em Advances in Mathematics} {\bf 263} (2014) 88--122.

\bibitem{armstrong2016rational}
D.~Armstrong, N.~A. Loehr, and G.~S. Warrington, {\it Rational parking
  functions and catalan numbers},  {\em Annals of Combinatorics} {\bf 20}
  (2016) 21--58.

\bibitem{gorsky2016affine}
E.~Gorsky, M.~Mazin, and M.~Vazirani, {\it Affine permutations and rational
  slope parking functions},  {\em Transactions of the American Mathematical
  Society} {\bf 368} (2016), no.~12 8403--8445.

\bibitem{kostka}
{SageMath}, ``{Computing Kostka number}.'' \url
  {https://doc.sagemath.org/html/en/reference/combinat/sage/combinat/sf/kfpoly.html},
  2025.
\newblock [Accessed: 15 September 2025].

\bibitem{hausel2019arithmetic}
T.~Hausel, M.~Mereb, and M.~L. Wong, {\it Arithmetic and representation theory
  of wild character varieties},  {\em Journal of the European Mathematical
  Society} {\bf 21} (2019), no.~10 2995--3052.

\bibitem{leven2014two}
E.~Leven, {\it Two special cases of the rational shuffle conjecture},  {\em
  Discrete Mathematics \& Theoretical Computer Science} (2014),
  no.~Proceedings.

\bibitem{qiu2020schur}
D.~Qiu and J.~Remmel, {\it Schur function expansions and the rational shuffle
  theorem},  {\em Journal of Combinatorial Theory, Series A} {\bf 175} (2020)
  105272.

\bibitem{MR3456698}
T.~Arakawa, {\it Associated varieties of modules over {K}ac-{M}oody algebras
  and {$C_2$}-cofiniteness of {$W$}-algebras},  {\em Int. Math. Res. Not. IMRN}
  (2015), no.~22 11605--11666.

\bibitem{Xie:2019vzr}
D.~Xie and W.~Yan, {\it {4d $\mathcal{N}=2$ SCFTs and lisse W-algebras}},  {\em
  JHEP} {\bf 04} (2021) 271, [\href{http://xxx.lanl.gov/abs/1910.0228}{{\tt
  arXiv:1910.0228}}].

\bibitem{sommers1997family}
E.~Sommers, {\it A family of affine weyl group representations},  {\em
  Transformation Groups} {\bf 2} (1997), no.~4 375--390.

\bibitem{thiel2016anderson}
M.~Thiel, {\it From anderson to zeta},  {\em Advances in Applied Mathematics}
  {\bf 81} (2016) 156--201.

\bibitem{haiman1994conjectures}
M.~D. Haiman, {\it Conjectures on the quotient ring by diagonal invariants},
  {\em Journal of Algebraic Combinatorics} {\bf 3} (1994), no.~1 17--76.

\bibitem{Song:2017oew}
J.~Song, D.~Xie, and W.~Yan, {\it {Vertex operator algebras of Argyres-Douglas
  theories from M5-branes}},  {\em JHEP} {\bf 12} (2017) 123,
  [\href{http://xxx.lanl.gov/abs/1706.0160}{{\tt arXiv:1706.0160}}].

\bibitem{szabo2021perversity}
S.~Szab{\'o}, {\it Perversity equals weight for painlev{\'e} spaces},  {\em
  Advances in Mathematics} {\bf 383} (2021) 107667.

\bibitem{berest2003finite}
Y.~Berest, P.~Etingof, and V.~Ginzburg, {\it Finite-dimensional representations
  of rational cherednik algebras},  {\em International Mathematics Research
  Notices} {\bf 2003} (2003), no.~19 1053--1088.

\bibitem{haglund2008q}
J.~Haglund, {\em The $ q, t $-Catalan numbers and the space of diagonal
  harmonics: with an appendix on the combinatorics of Macdonald polynomials},
  vol.~41.
\newblock American Mathematical Soc., 2008.

\end{thebibliography}\endgroup

\end{document}